\documentclass{eas}
\usepackage{graphicx}
\usepackage{color}
\usepackage{hyperref}

\begin{document}

%%-----------------------------
%%      the top matter
%%-----------------------------
\title{Stellar Winds and Planetary Atmospheres}
\author{Colin P. Johnstone} \address{Natural History Museum, Burgring 7, A-1010 Vienna, Austria} \secondaddress{University of Vienna, Department of Astrophysics, T\"{u}rkenschanzstrasse 17, 1180 Vienna, Austria}

\begin{abstract}
Interactions between the winds of stars and the magnetospheres and atmospheres of planets involve many processes, including the acceleration of particles, heating of upper atmospheres, and a diverse range of atmospheric loss processes.
Winds remove angular momentum from their host stars causing rotational spin-down and a decay in magnetic activity, which protects atmospheres from erosion.
While wind interactions are strongly influenced by the X-ray and ultraviolet activity of the star and the chemical composition of the atmosphere, the role of planetary magnetic fields is unclear.
In this chapter, I review our knowledge of the properties and evolution of stellar activity and winds and discuss the influences of these processes on the long term evolution of planetary atmospheres.
I do not consider the large number of important processes taking place at the surfaces of planets that cause exchanges between the atmosphere and the planet's interior.  
\end{abstract}
\maketitle
%%-----------------------------
%%      your text
%%-----------------------------

\section{Introduction}

The discoveries in the last decades of thousands of planets outside our solar system has led to questions of planetary habitability becoming a central theme in modern astrophysics. 
Central to these questions is the question of how planetary atmospheres evolve, and what conditions are necessary to form an atmosphere that can sustain life (for a review, see G\"udel et al. \cite{Guedel14}).
The formation and subsequent evolution of a planet's atmosphere are determined by exchanges of volatiles between the planet's interior and atmosphere and by losses of gas from the system to space by a  large number of processes.
These loss processes take place mostly in response to magnetic activity related output of the planet's host star, and especially its emission of X-ray and ultraviolet radiation and its wind. 
In this chapter, I discuss the properties and evolution of stellar winds and the various ways in which they influence planetary atmospheres.
For a more comprehensive review on the topic, see the recent book by Linsky (\cite{Linsky19}).

Atmospheres form on planets in several ways and different processes lead to atmospheres composed of different mixtures of chemical species.
Planets that form to sufficient mass (typically above 0.1~M$_\oplus$) in the first few million years when a significant gas disk is still present can accumulate hydrogen and helium dominated primordial atmospheres and the amount of gas accumulated depends primarily on the mass of the planet at the end of the disk phase (St\"okl et al. \cite{Stoekl16}). 
Lammer et al. (\cite{Lammer14}) showed that since higher mass planets accumulate more atmosphere and lose them slower after the disk is gone, planets with masses above approximately an Earth mass keeps these atmospheres for their entire lives and a similar result was found for planets orbiting M~dwarfs by Owen \& Mohanty (\cite{OwenMohanty16}).
Observations of low-mass exoplanets with large radii (e.g. Lissauer et al. \cite{Lissauer11}; Rogers \cite{Rogers15}) suggest that many terrestrial planets have such envelopes, and especially higher mass planets (e.g. Rogers \cite{Rogers15}).
Evidence from the isotopic abundances of noble gases in the lower mantle suggest that the Earth did in fact possess a primordial atmosphere (Yokochi \& Marty \cite{YokochiMarty04}; Williams \& Mukhopadhyay \cite{WilliamsMuk19}), though disagreement exists about how to interpret these results (P\'eron et al. \cite{Peron16}). 
The isotopic compositions of hafnium and tungsten in the mantle suggest that the Earth likely grew to half its mass in approximately 10~Myr and its full mass within approximately 50~Myr (Kleine et al. \cite{Kleine09}).
Approximately half of all gas disks are thought to be gone by $\sim$2~Myr (Mamajek \cite{Mamajek09}), though there is evidence that a significant minority of stars retain their disks for much longer than 10~Myr (Pfalzner et al. \cite{Pfalzner14}).
It is plausible that the Earth gained a small primordial atmosphere, though since this atmosphere is no longer present, it likely did not form to its full mass during the gas disk phase.

Volatile species are distributed throughout several reservoirs in a planet, including the mantle, the crust, the ocean, and the atmosphere.
Impacts during a planet's growth phase lead to volatile species being released into the atmosphere from both the impactor and the planet, and what is not released from the impactor can be stored within the planet outer layers for release at later times (Elkins-Tanson \& Seager \cite{EklinsSeager08}).
Energy released from these impacts likely cause the planets to be so hot that a liquid magma ocean is present during and immediately after the main growth period.
As the planet solidifies and forms a mantle and crust, large amounts of volatiles can be degassed into the atmosphere (Elins-Tanton \cite{ElkinsTanton08}; Salvador et al. \cite{Salvador17}) and if it is not rapidly lost to space, this atmosphere can lengthen the magma ocean phase by slowing the radiative cooling of the surface (Nikolaou et al. \cite{Nikolaou19}).
In later phases, tectonic related activity, such as mantle outgassing from volcanoes, is important for forming and replenishing atmospheres (e.g. Noack et al. \cite{Noack14}), which can even be directly influenced by stellar magnetic fields (Kislyakova et al. \cite{Kislyakova17}; Kislyakova \& Noack \cite{KislyakovaNoack20}).
Surface processes can also transfer gas into the interior and this can be more important for removing atmospheres than escape to space.

\begin{figure}[!h]
\centering
\includegraphics[width=0.19\textwidth]{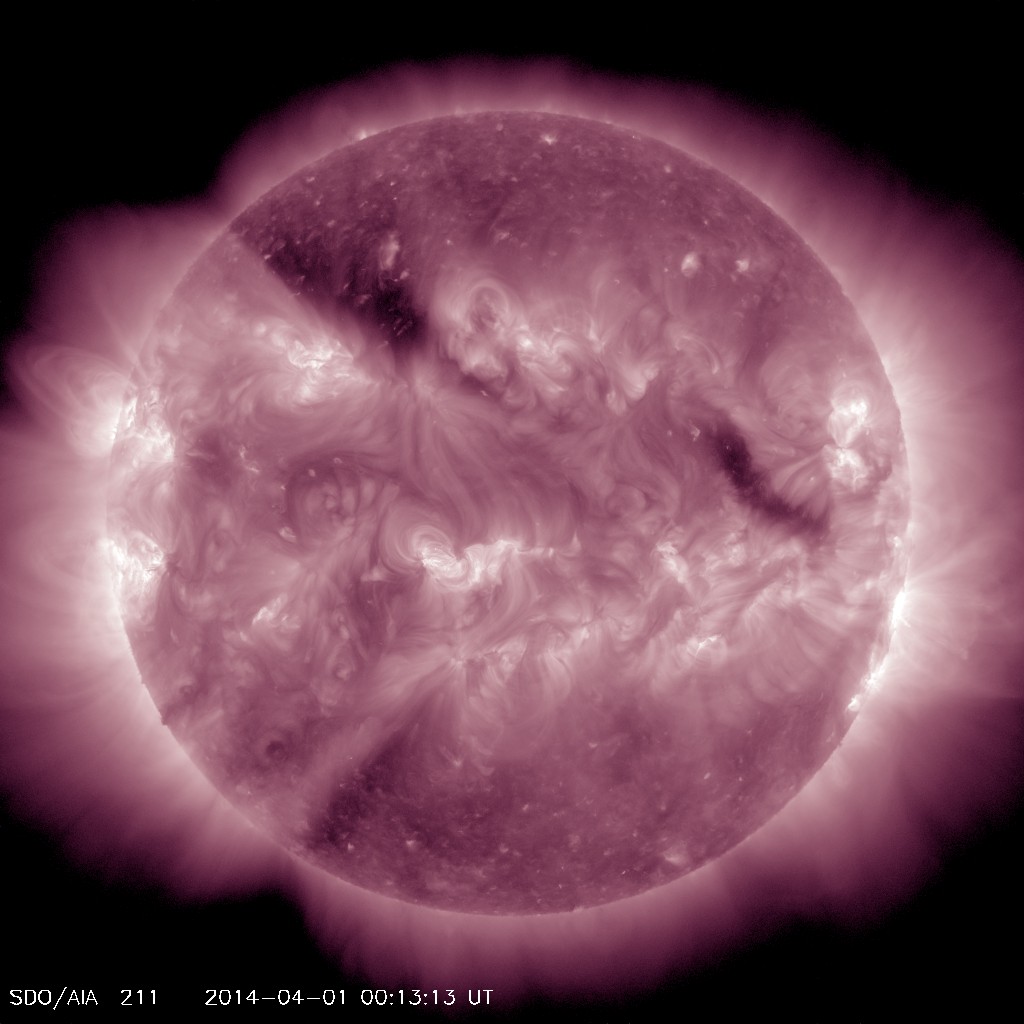}
\includegraphics[width=0.19\textwidth]{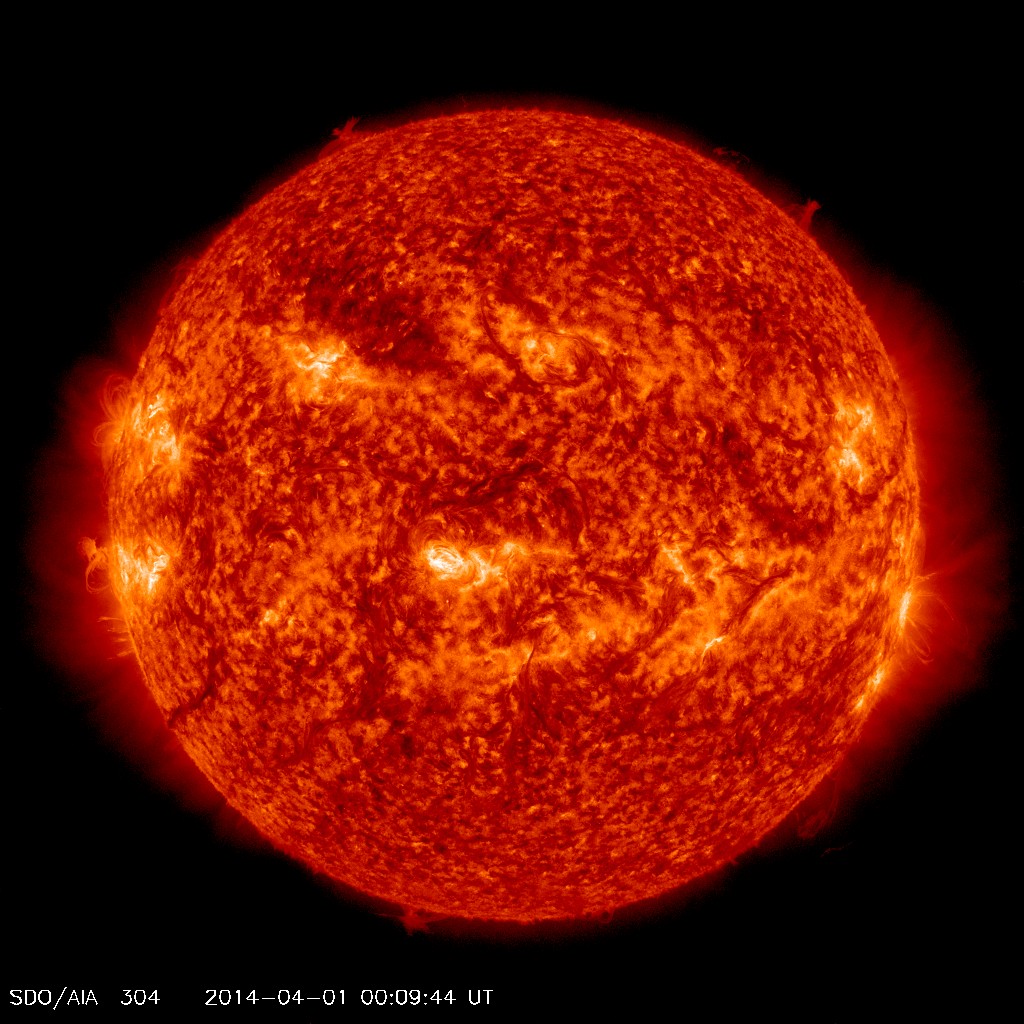}
\includegraphics[width=0.19\textwidth]{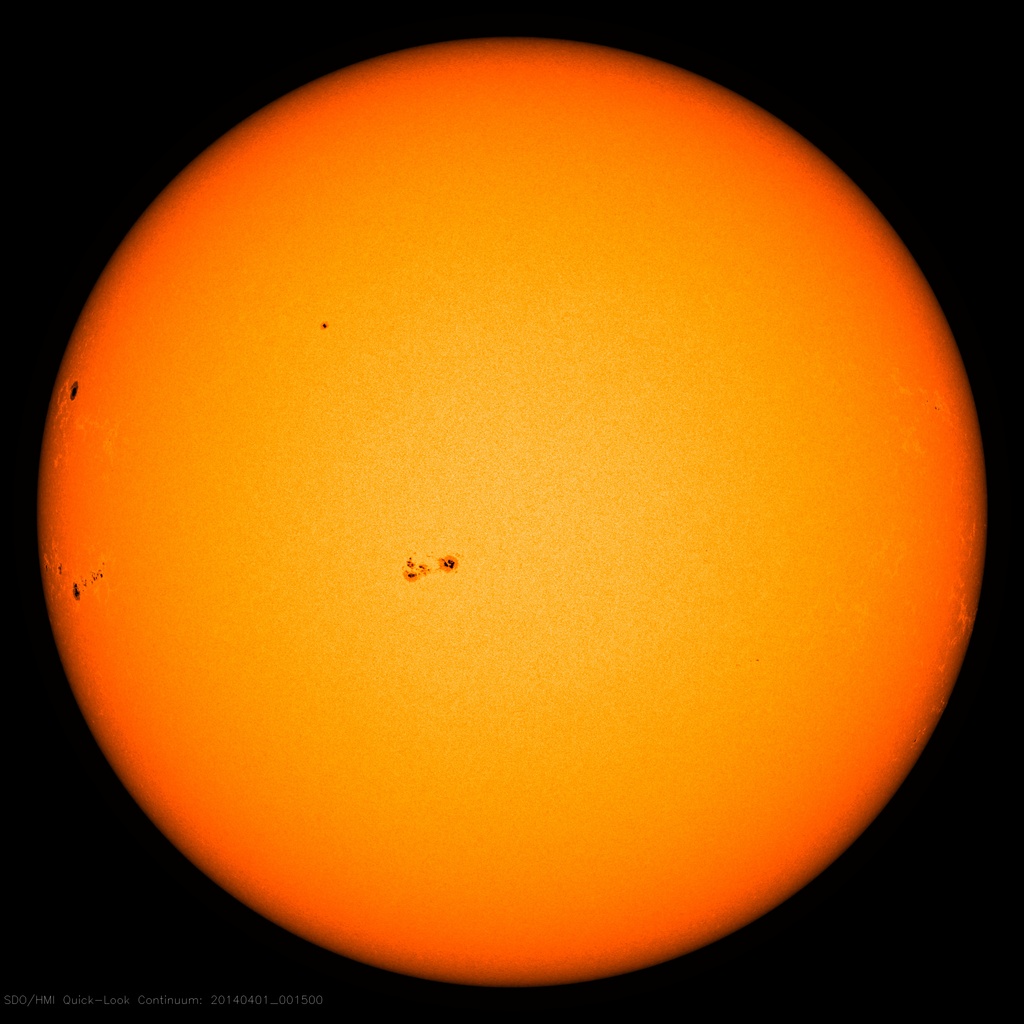}
\includegraphics[width=0.19\textwidth]{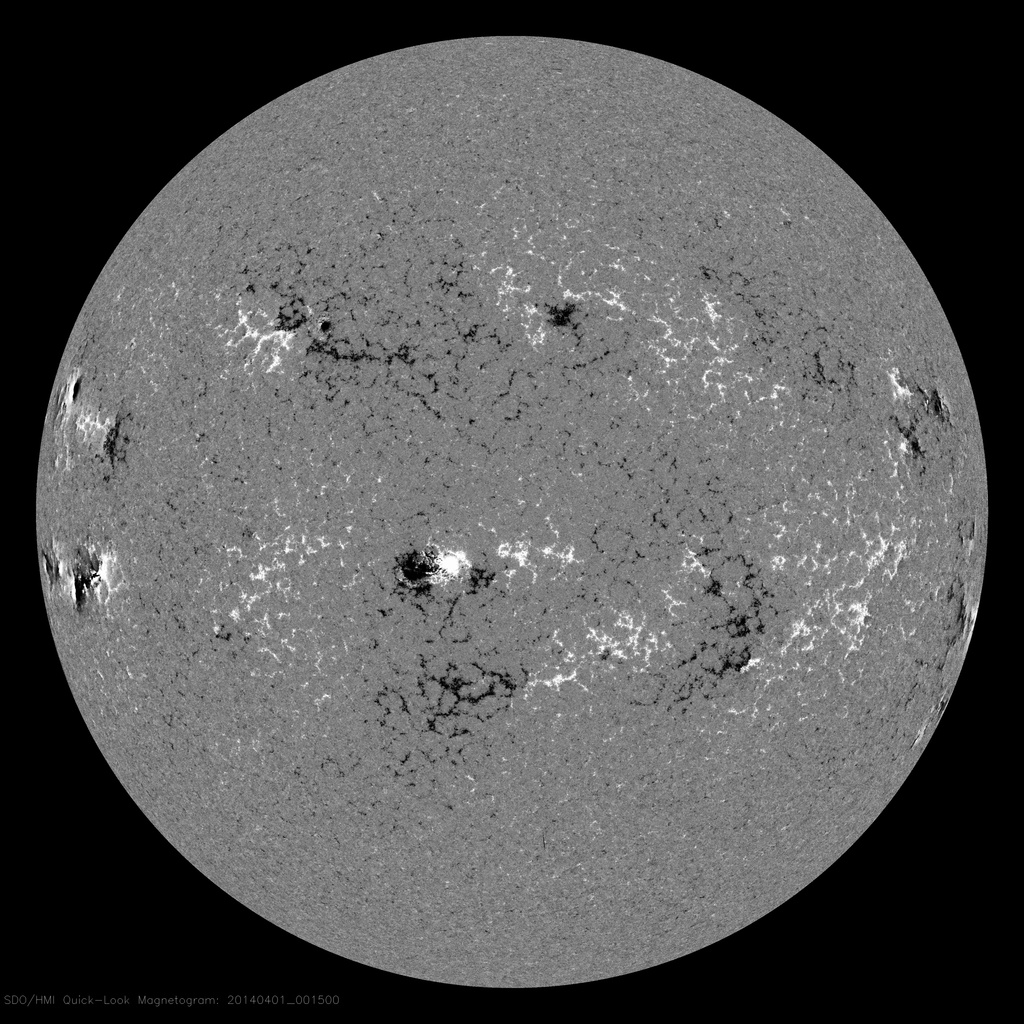}
\includegraphics[width=0.76\textwidth]{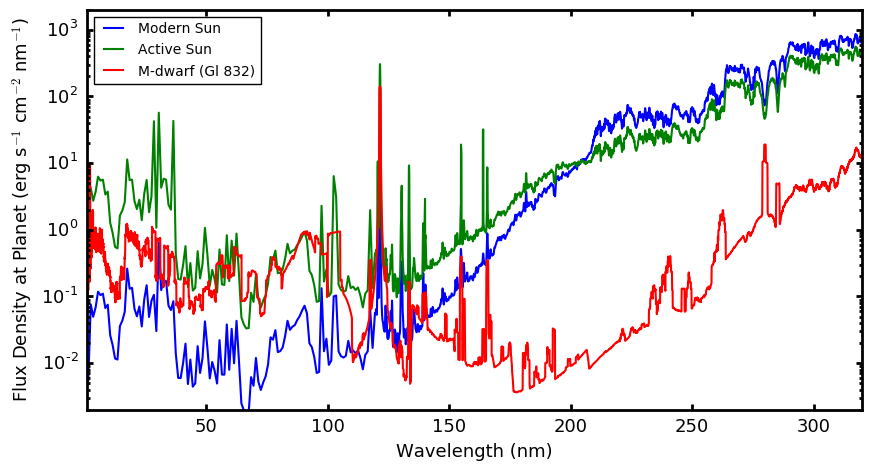}
\includegraphics[width=0.76\textwidth]{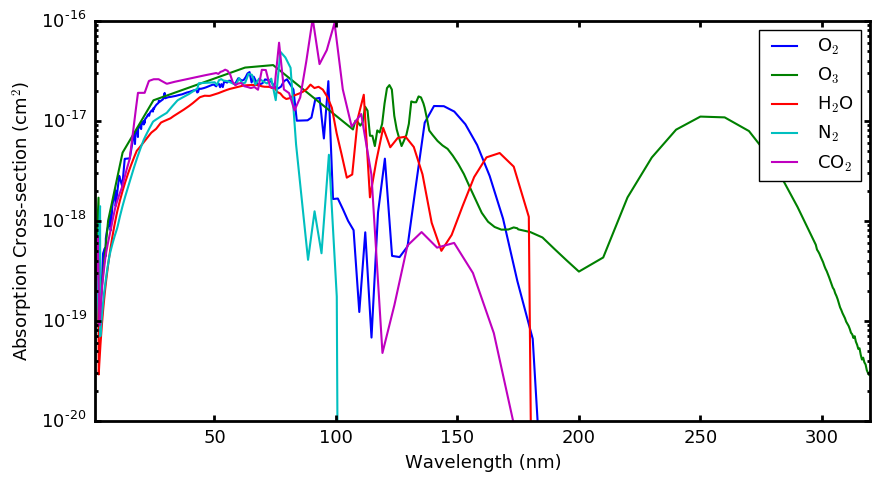}
\vspace{-5mm}
\caption{
\emph{
Upper panels: 
images of the Sun from the Solar Dynamics Observatory courtesy of NASA/SDO and the AIA, EVE, and HMI science teams.
From left to right, the panels show the corona at \mbox{$2 \times 10^6$~K}, the chromosphere and transition region at \mbox{$5 \times 10^4$~K}, the photosphere, and the surface magnetic field structure (with black and white showing opposite polarity magnetic field).
Middle panel: 
typical XUV spectra of the modern Sun during activity maximum (blue line) and a more active young Sun (green line) from Claire et al. (\cite{Claire12}), and the low-mass M-dwarf Gliese~832 as modelled by Fontenla et al. (\cite{Fontenla16}).
The solar spectra are scaled to 1~AU and the Gliese-832 spectrum is scaled to 0.162~AU, which is the orbit of a known habitable zone planet.
The lower flux of the Gliese~832 spectrum longward of $\sim$150~nm is due to the much cooler photosphere of this star, meaning its photospheric spectrum is shifted to longer wavelengths.
Note that the resolution of the Gliese-832 spectrum has been reduced to be similar to the other two spectra. 
Lower panel: 
XUV absorption cross-sections for several important species in planetary atmospheres from the PHIDRATES database (Huebner \& Mukherjee \cite{HuebnerMukherjee15}). 
}
}
\vspace{-4mm}
\label{fig:xuvspectrum}
\end{figure}

The fact that losses to space can play a major role in an atmosphere's evolution can be seen in our own solar system.
Venus for example has likely undergone massive loss of water due to the dissociation and loss of water molecules in the upper atmosphere (Chassefi{\`e}re et al. \cite{Chassefiere12}).
Since hydrogen atoms are lost more efficiently than the heavier deuterium atoms, this massive water loss can be seen in Venus' atmosphere by the very large D/H ratio compared to other solar system bodies (Marcq et al. \cite{Marcq18}).
While Mars currently has a very thin atmosphere, there is strong evidence that it previously had liquid water on its surface (Kite \cite{Kite19}) meaning that it must have had a much thicker atmosphere that has since been lost.
This is consistent also with the atmosphere’s isotopic composition which contains enhanced abundances of heavier isotopes of certain elements relative to their lighter counterparts (Jakosky et al. \cite{Jakosky17}). 
Outside our solar system, atmospheric escape has been directly observed from exoplanets (Ehrenreich et al. \cite{Ehrenreich15}) and this escape likely has observable influences on the distributions of exoplanetary radii as measured by spacecraft such as \emph{Kepler}. 
For example, the `evaporation valley' has been predicted to result from such losses and has significant observational support (Owen \& Wu \cite{OwenWu13}, \cite{OwenWu17}; Fulton et al. \cite{Fulton17}).

A star's magnetic field is one of the most important factors for the structure of its outer atmosphere and causes the surface of the star to be inhomogeneous and time variable, which can complicate the detection and characterisation of exoplanets (Collier~Cameron \cite{Cameron18}). 
Field lines anchored in the photosphere where the plasma beta (the ratio of the thermal pressure to the magnetic pressure) is much greater than unity are moved around by convective motions of the plasma.
Although it is not yet clear what the dominant processes are, energy from these convective motions is transferred by the magnetic field to higher altitudes where it heats the gas to several tens of thousands of K in the chromosphere and to MK temperatures in the corona.
The corona is the large outer atmosphere of the star and has a plasma beta much less than unity, meaning that the dynamics of the plasma is controlled by the magnetic field.
This results in the emission of X-ray and extreme ultraviolet (EUV) radiation from the corona, and EUV and far ultravolet (FUV) radiation from the chromosphere.
The term `XUV' is commonly used to refer to a star's X-ray and ultraviolet spectrum, though definitions of this term vary and often includes only X-ray and EUV radiation; in this review, I do not attach an exact definition to the term unless specified. 
Another effect of these high temperatures is the acceleration of the plasma away from the star in the form of a transonic (starting subsonic and accelerating to supersonic speeds) wind. 
The outer layers of the Sun's atmosphere and the resulting X-ray and ultraviolet spectrum of the modern Sun, a younger more active Sun, and a low mass M dwarf, are shown in Fig.~\ref{fig:xuvspectrum}.
This high energy radiation is important since most common atmospheric species absorb radiation at these wavelengths very efficiently, meaning that the radiation is absorbed high in the atmospheres of planets where the gas densities are low.
This heats and expands the upper atmosphere, making it more susceptible to interactions with the star's wind.

\section{Stellar rotation and activity evolution}

It is important to consider the long term evolution of stellar rotation since we should expect that this will drive a corresponding evolution in stellar winds and their interactions with atmospheres.
Possibly the most important way that winds influence the atmospheres of planets is by removing angular momentum from their host stars, causing a decay in both their rotation rates and their emission of X-ray and extreme ultraviolet radiation.
In the last two decades, our observational knowledge of rotational evolution has improved dramatically, especially due to rotation distributions in young stellar clusters being determined by ground-based photometric monitoring campaigns (e.g. Hartman et al. \cite{Hartman10}) and more recently by the \emph{K2} mission (e.g. Rebull et al. \cite{Rebull16}). 
This has driven the development of more detailed physical models and a comprehensive though still incomplete understanding of how the rotation rates of stars evolve from their initial births to the end of the main-sequence.

\subsection{The observational picture}

\begin{figure}
\centering
\includegraphics[width=0.9\textwidth]{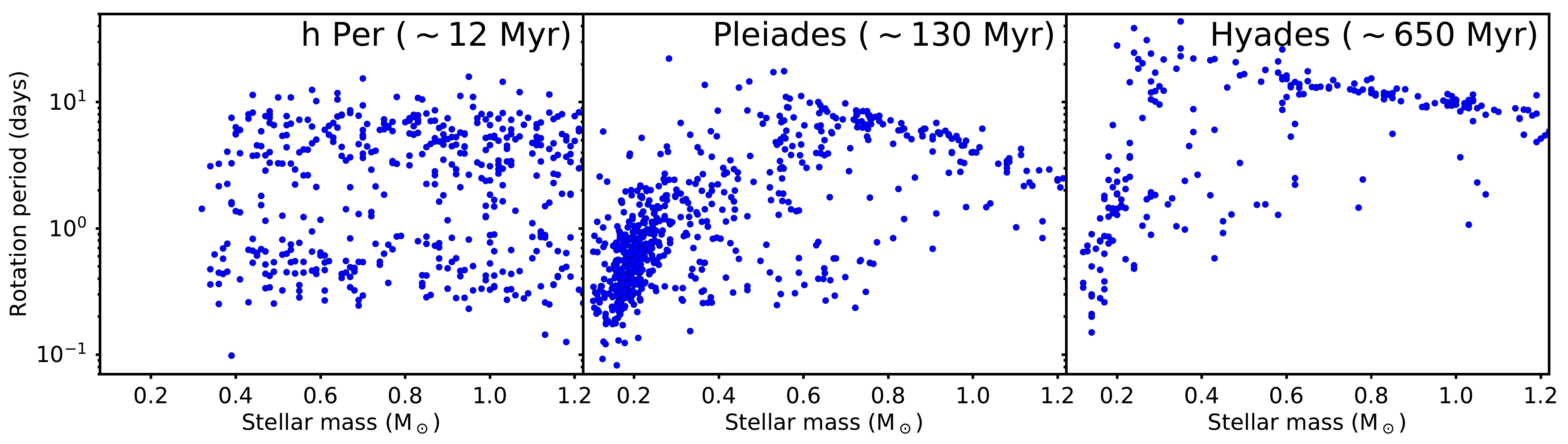}
\includegraphics[width=0.9\textwidth]{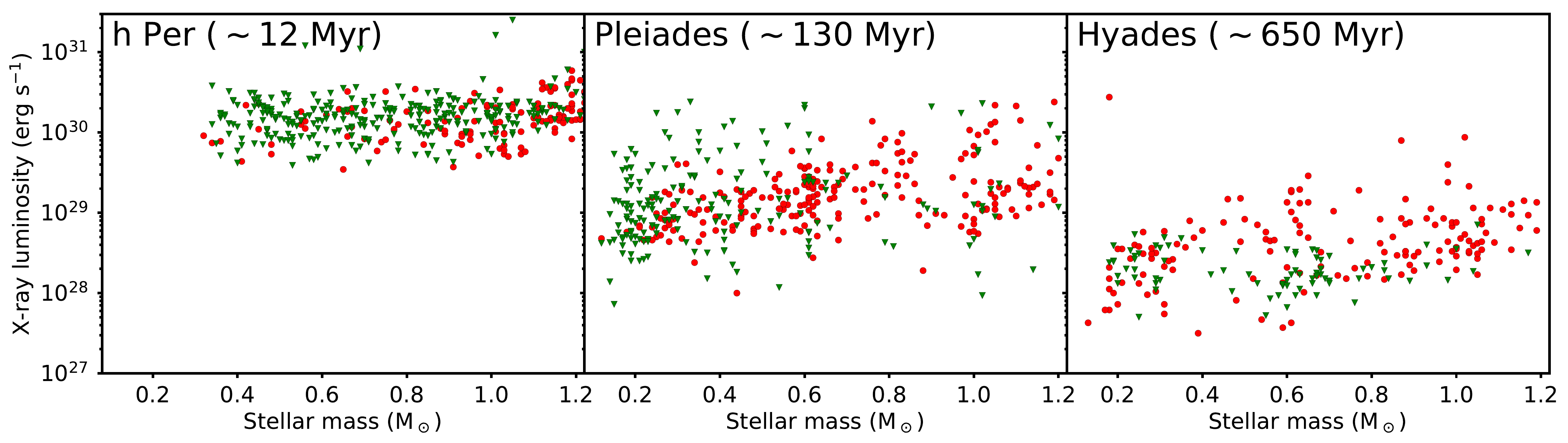}
\vspace{-5mm}
\caption{
\emph{
Rotation period and X-ray luminosity distributions in the young clusters h~Per, Pleiades, and Hyades.
Rotation periods are from Moraux et al. (\cite{Moraux13}) for h~Per, Hartman et al. (\cite{Hartman10}) and Rebull et al. (\cite{Rebull16}) for Pleiades, and Douglas et al. (\cite{Douglas16}) for Hyades.
The X-ray luminosities are from Argiroffi et al. (2016) for h~Per, and N{\'u}{\~n}ez et al. (2016) for both Pleiades and Hyades.
}
}
\vspace{-4mm}
\label{fig:rotxraydists}
\end{figure}

Stars in young clusters with ages of less than 5~Myr have a range of rotation rates distributed between approximately 1 and 50 times the solar rotation rate (Affer et al. \cite{Affer13}).
This distribution appears to be initially mass independent and during the first few Myr, it remains approximately constant in time (Rebull et al. \cite{Rebull04}), though there is evidence that stars with masses below approximately 0.5~M$_\odot$ might spin up during this time (Henderson \& Stassun \cite{HendersonStassun12}).
At an age of 12~Myr, the rotation distribution in young cluster h~Per studied by Moraux et al. (\cite{Moraux13}) shows two changes relative to younger clusters. Firstly, the distribution has been shifted to more rapid rotation, especially for the rapidly rotators.
Secondly, the distribution has separated into a rapidly rotating track and a slowly rotating track, with few stars in between.
This distribution can be seen in Fig.~\ref{fig:rotxraydists} and the evolution of this distribution on the main-sequence can be seen by comparison with the Pleiades ($\sim$130~Myr) and Hyades ($\sim$650~Myr) clusters.
For Pleides, rotation rates from Hartman et al. (\cite{Hartman10}) and Rebull et al. (\cite{Rebull16}) give us a detailed description of the rotation at this important age. 
At this age, most stars with masses above approximately 0.5~M$_\odot$ are part of the slowly rotating track, but a tail of rapid rotators is also present, while most lower mass stars remain rapidly rotating.
By 650~Myr, this distribution has shifted to slower rotation and more stars have converged onto the slowly rotating track, though this convergence is slower for lower mass stars.
This spin-down and convergence continues throughout their main-sequence evolution, though recent measurements of K dwarfs seem to show a temporary halt in their spin-down between ages of approximately 500-1000 Myr (Curtis et al. \cite{Curtis19}).
Although at young ages they remain rapidly rotating longer, at later ages lower mass stars are more slowly rotating than higher mass stars (Nielsen et al. \cite{Nielsen13}).
Irwin et al. (\cite{Irwin11}) provided evidence that some fully convective M~dwarfs do not converge onto the slowly rotating track and remain rapidly rotating for timescales longer than 10~Gyr.

\begin{figure}
\centering
\includegraphics[width=0.95\textwidth]{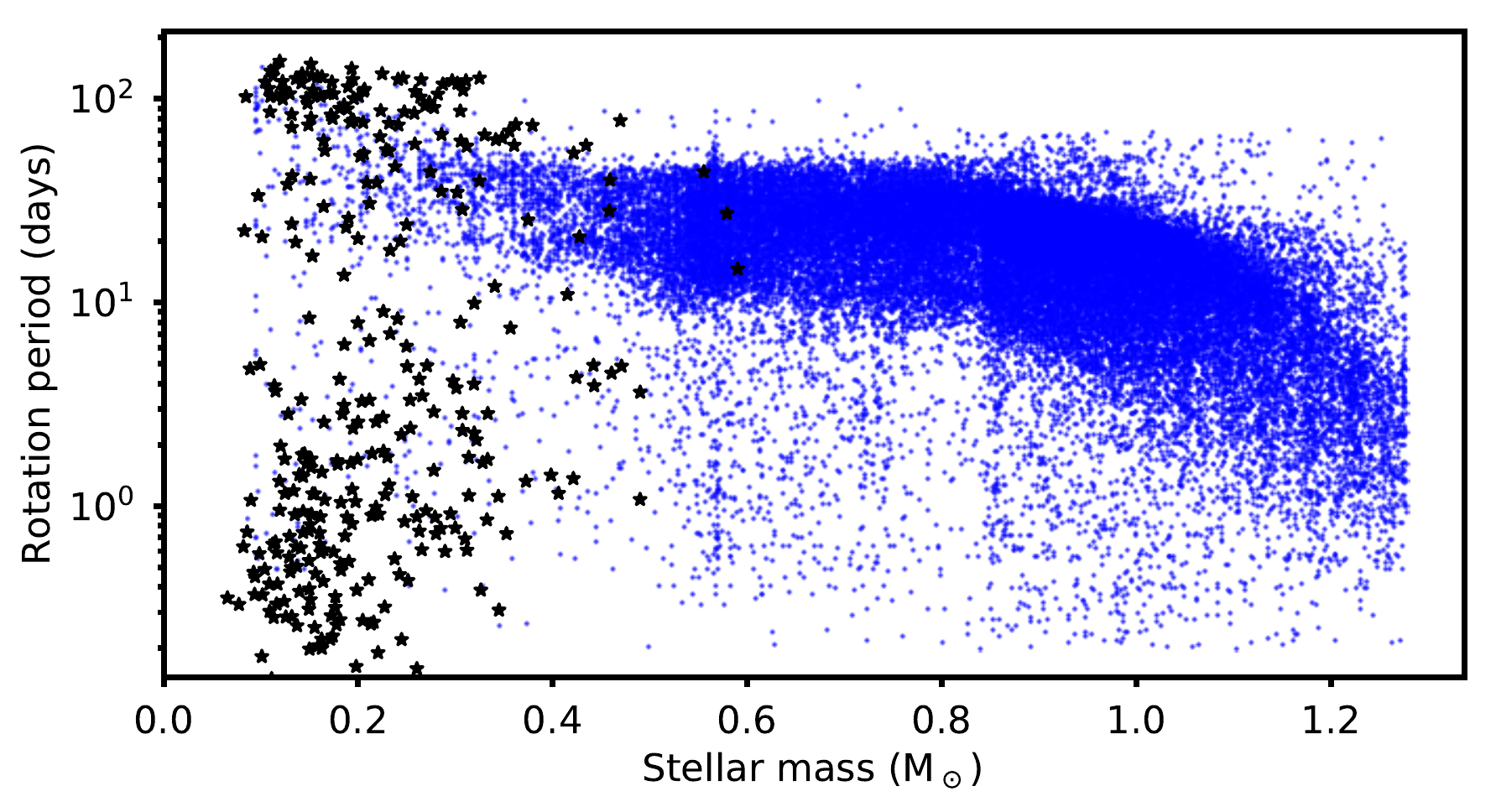}
\vspace{-5mm}
\caption{
\emph{
Rotation distribution for approximately 40,000 field stars from McQuillan et al. (\cite{McQuillan14}) and Santos et al. (\cite{Santos19}) measured by Kepler (blue points) and for low mass field stars from Newton et al. (\cite{Newton16}) by the MEarth Project (black stars).
}
}
\vspace{-4mm}
\label{fig:laterrot}
\end{figure}

While observations constrain the first Gyr of rotational evolution, how stars spin down after this time is less well constrained.
Especially due to \emph{Kepler} measurements, we know the rotation rates of a very large number of field stars (McQuillan et al. \cite{McQuillan14}; Santos et al. \cite{Santos19}), which are informative but of more limited use since the ages of these stars are difficult to determine (Soderblom et al. \cite{Soderblom14}). 
In Fig.~\ref{fig:laterrot}, I show the rotation-mass distribution for field stars measured by the \emph{Kepler} spacecraft and for lower mass stars by Newton et al. (\cite{Newton16}).
At later ages, higher mass stars tend to be rotating more rapidly, which could be surprising given that in the first billion years, lower mass stars remain rapidly rotating longer. 
It is also shown from the MEarth Project sample of lower mass stars that even for field stars, most of which likely have ages of a few Gyr or more, there are still a large number of very rapid rotators.

The rotational evolution of solar mass stars was described in by Skumanich (\cite{Skumanich72}) by \mbox{$P_\mathrm{rot} \propto t^{0.5}$}, where $P_\mathrm{rot}$ is the rotation period and $t$ is the age.
While this was originally based only on comparing average equatorial rotation velocities from Hyades and Pleiades with the rotation of the Sun, this relation has been shown to be approximately accurate for solar mass stars at later ages, though this conclusion is dependent on the assumption that the Sun's modern rotation is a good proxy for the rotation rates of solar mass stars with similar ages.
The late mass dependence for rotationat later ages is more difficult to constrain and it is typical to assume that $P_\mathrm{rot}$ depends independently on mass and age and to constrain the mass dependence using the shape of the slow rotating track in young clusters (e.g. Barnes \cite{Barnes07}; Mamajek \& Hillenbrand \cite{MamajekHillenbrand08}).
The reliability of this assumption is difficult to test given the lack of reliable ages for older stars with masses below that of the Sun.
Based on this assumption, Mamajek \& Hillenbrand (\cite{MamajekHillenbrand08}) derived
\begin{equation} \label{eqn:gyroprot}
P_\mathrm{rot} = a \left[ (B-V)_0 - c \right]^b t^n
\end{equation}
where \mbox{$a=0.407$}, \mbox{$b=0.325$}, \mbox{$c=0.495$}, \mbox{$n=0.566$}, $t$ is in Myr, and $P_\mathrm{rot}$ is in days.
Here, \mbox{$(B-V)_0$} represents stellar mass and solar mass main-sequence stars have values of approximately 0.66, whereas 0.5 and 0.1~M$_\odot$ stars have values of approximately 1.5 and 2.0 respectively (Pecaut \& Mamajek \cite{PecautMamajek13}).

\subsection{The epochs of rotational evolution} \label{sect:rotepochs}

\begin{figure}
\centering
\includegraphics[width=0.49\textwidth]{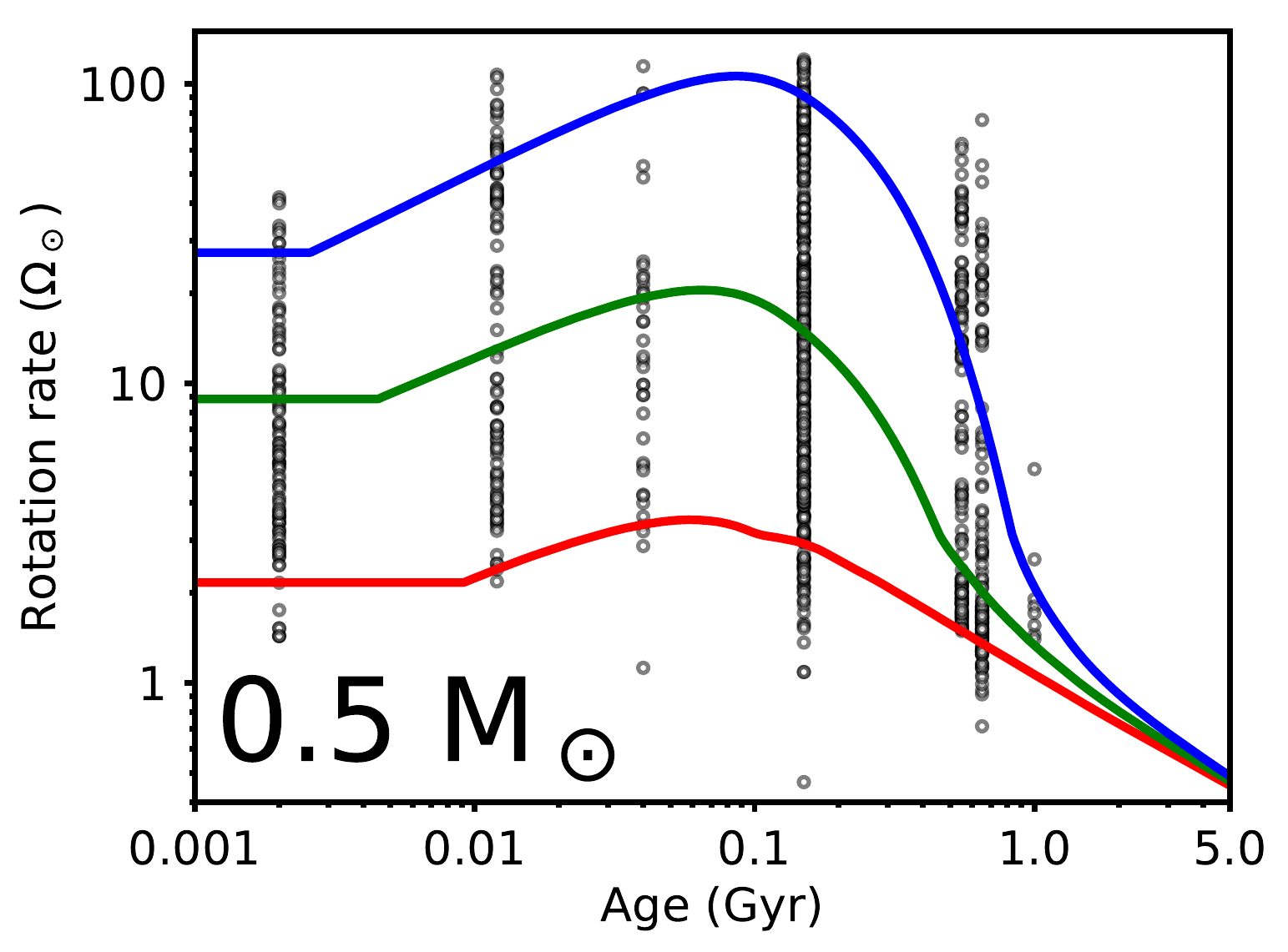}
\includegraphics[width=0.49\textwidth]{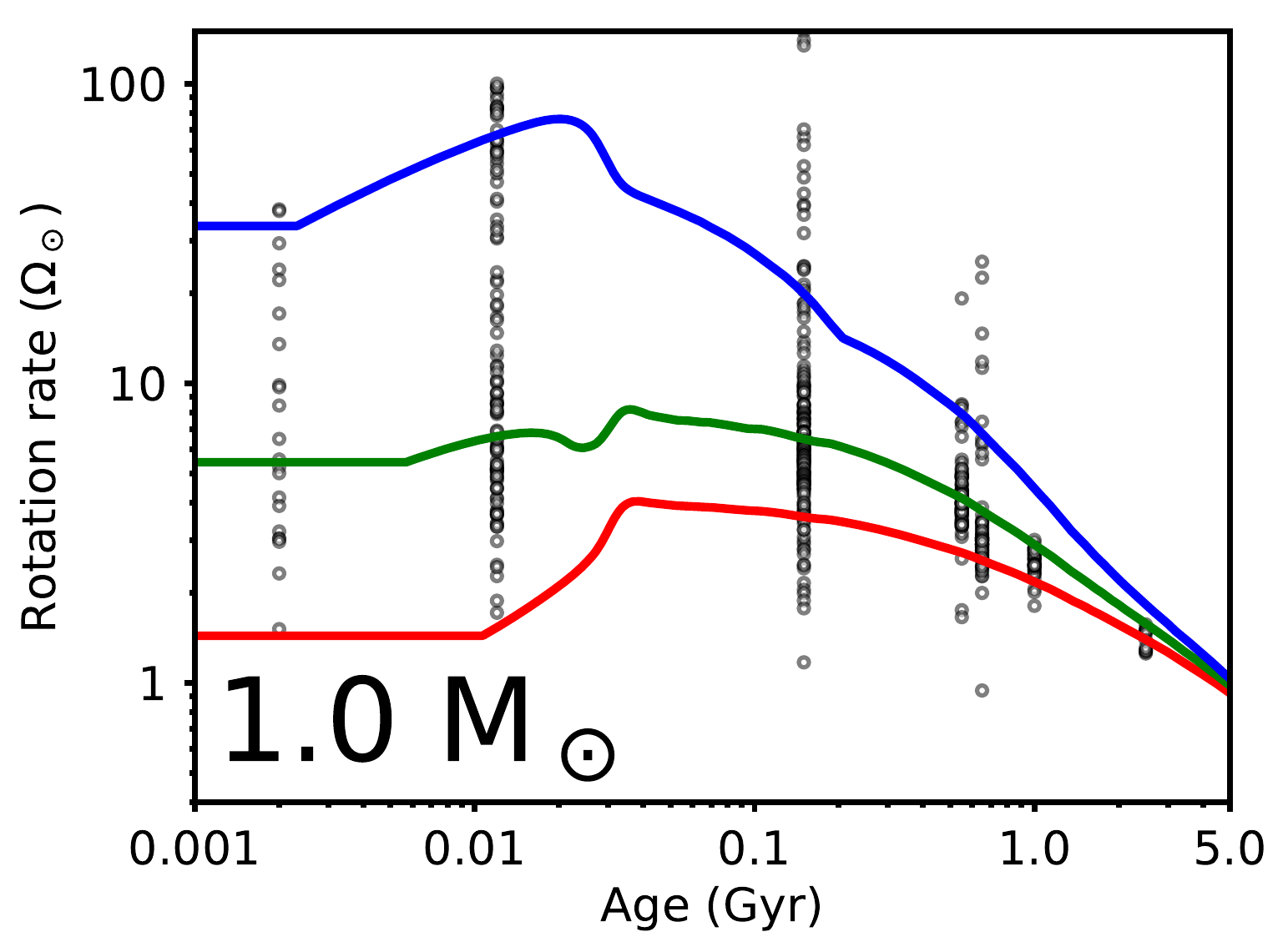}
\vspace{-3mm}
\caption{
\emph{
Evolutionary tracks for surface rotation of stars with masses of 0.5 and 1.0~M$_\odot$ with different initial rotation rates.
In both panels, the red, green, and blue lines show stars at the 5$^\mathrm{th}$, 50$^\mathrm{th}$, and 95$^\mathrm{th}$ percentiles of the observed rotation distribution at 150~Myr and the black circles show observed rotation rates from several young clusters.
The rotation tracks are calculated using the model developed by Johnstone et al. (\cite{Johnstone19b}) and extended by Johnstone et al. (submitted).
}
}
\vspace{-4mm}
\label{fig:rottracks}
\end{figure}

A star's long term rotational evolution depends primarily on its mass and its early rotation rate (Gallet \& Bouvier \cite{GalletBouvier15}).
The latter is important because stars that are born as rapid rotators remain rapidly rotating much longer than those born as slow rotators. 
Other important factors include the star's metallicity (Amard \& Matt \cite{AmardMatt20}) and possibly the lifetime of its circumstellar disks during its classical T Tauri phase  (Herbst et al. \cite{Herbst02}).
A star's rotational evolution from 1~Myr to the end of the main-sequence can be broken down into several important epochs: these are the phases of disk-locking, pre-main-sequence spin-up, and main-sequence spin-down.
The rotational evolution of 0.5 and 1~M$_\odot$ stars is shown in Fig.~\ref{fig:rottracks}.

The observed lack of rotational spin up for classical T Tauri stars in the first few million years is surprising since they are contracting on the pre-main-sequence and are accreting high angular momentum material from their disks.
This means that angular momentum must be removed from the stars and potentially also from the accreting material during this time, though which mechanisms are responsible are poorly understood. 
A possibility is that very strong winds originating from the surfaces of the stars exert a strong enough spin-down torque to keep the stars in rotational equilibrium (Matt \& Pudritz \cite{MattPudritz08a}) and this strong wind has been suggested to be powered by accretion onto the surface from the gas disk (Cranmer \cite{Cranmer08}; Cranmer \cite{Cranmer09a}).
This phase is commonly referred to as `disk-locking' given the assumption that interactions with the disk are responsible (for a review, see Bouvier et al. (\cite{Bouvier14}) and it is generally assumed that this phase ends around the time that these disks disperse, which is typically within the first 10~Myr (Mamajek \cite{Mamajek09}), after which stars spin up due to pre-main-sequence contraction.
This spin-up is reduced by the removal of angular momentum by stellar winds and stops approximately at the zero-age main-sequence (ZAMS).

After the ZAMS, the main process is wind driven spin-down.
Several factors are important for how rapidly winds remove angular momentum from their host stars and these are mostly related to the star's mass, radius, rotation rate, and the properties of the wind and global magnetic field (Kawaler \cite{Kawaler88}; Matt \& Pudritz \cite{MattPudritz08b}).
Characterising the dependence of angular momentum loss on these parameters is difficult and recent magnetohydrodynamic models have been used to explore this question (e.g. Matt et al. \cite{Matt12}).
Important is not only the strength of the magnetic field, but also its geometry (Garraffo et al. \cite{Garraffo15}; Finley et al. \cite{Finley18}) and thermal structure (Cohen \cite{Cohen17}; Pantolmos \& Matt \cite{Pantolmos17}).

As I discuss in the next section, stellar activity has a saturated (activity independent of rotation) regime and an unsaturated (activity dependent of rotation) regime.
Observationally, it appears that main-sequence stars in the saturated regime show exponential spin-down which implies that \mbox{$\dot{\Omega} \propto \Omega$}, where \mbox{$\dot{\Omega} = d\Omega / dt$}, whereas the Skumanich spin-down of unsaturated stars implies \mbox{$\dot{\Omega} \propto \Omega^3$}.
This difference can be easily understood if mass loss rates and magnetic field strengths are independent of rotation in the saturated regime and depend strongly on rotation in the unsaturated regime. 
In both regimes, the positive dependence of \mbox{$\dot{\Omega}$} on $\Omega$ means that the wide distribution of rotation rates seen at the start of the main-sequence converges until all stars with the same mass follow a single spin-down track. 
Activity saturation also explains why rapidly rotating lower mass stars spin down slower than rapidly rotating higher mass stars, despite the fact that at later ages, lower mass stars spin down more rapidly.
Since lower mass stars saturate at lower rotation rates, in the saturated regime they feel a reduced spin-down wind torque.
In the unsaturated regime, wind torques are approximately mass independent (Gallet \& Bouvier \cite{GalletBouvier15}; Johnstone et al. \cite{Johnstone15b}), meaning that lower mass stars spin down more rapidly due to their smaller moments of inertia. 

\begin{figure}
\centering
\includegraphics[width=0.99\textwidth]{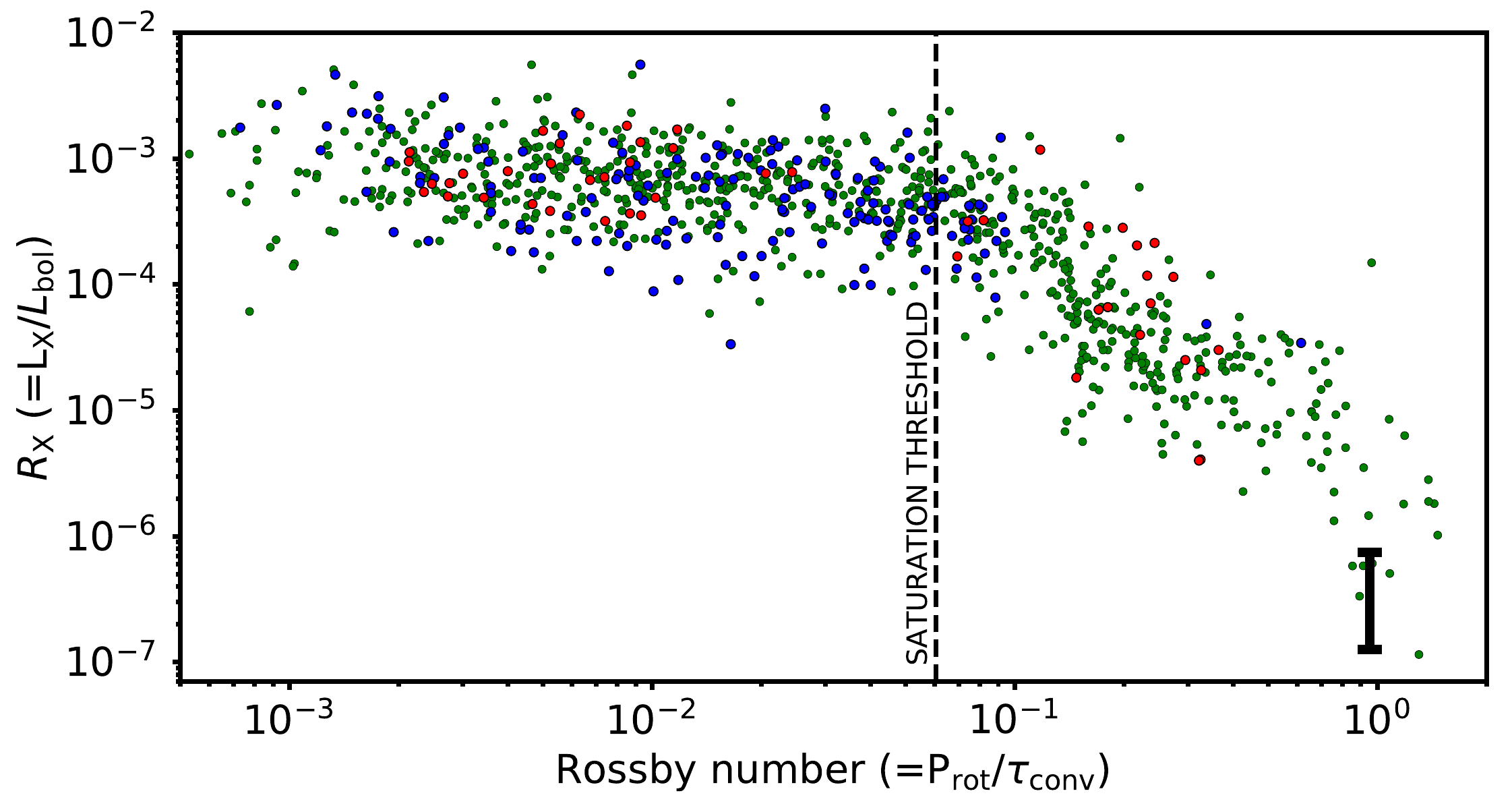}
\vspace{-3mm}
\caption{
\emph{
The correlation between stellar X-ray emission and rotation characterised as a $Ro$--$R_\mathrm{X}$ relation. 
Here, green circles show main-sequence stars from Wright et al. (\cite{Wright11}), blue circles show stars in the young 12~Myr old cluster h~Per from Argiroffi et al. (\cite{Argiroffi16}), and red circles show fully convective main-sequence M dwarfs from Jeffries et al. (\cite{Jeffries11}) and Wright et al. (\cite{Wright18}).
I do not include upper limits from these studies and for main-sequence M~dwarfs these upper limits could indicate that the slope in the unsaturated regime is steeper than for higher mass stars.
The black line on the lower right shows the range of values for the modern Sun (specifically the 10$^\mathrm{th}$ to the 90$^\mathrm{th}$ percentiles) calculated from the Flare Irradiance Spectrum Model (Chamberlin et al. \cite{Chamberlin07}).
All values of $Ro$ have been calculated using the convective turnover times from Spada et al. (\cite{Spada13}).
}
}
\vspace{-4mm}
\label{fig:RoRx}
\end{figure}

\subsection{Activity and rotation} \label{sect:rotactivity}

Empirically, it is known that the activity of a star depends primarily on its mass, age, and rotation rate.
This dependence on rotation can be seen in direct measurements of photospheric magnetic fields (Vidotto et al. \cite{Vidotto14}), in chromospheric activity indications such as Ca II H\&K and H$\alpha$ lines (Boro Saikia et al.~\cite{BoroSaikia18}), and in coronal X-ray emission (Pizzolato et al. \cite{Pizzolato03}).
For most activity indicators, the rotation relation has two regimes.
At slow rotation a strong rotation--activity relation is seen with more rapidly rotating stars being more active and this regime is called the unsaturated regime.
At fast rotation, the rotation--activity relation is very flat with almost no difference between stars with different rotation rates and this regime is called the saturated regime. 
On the pre-main-sequence, the rotation rate of the saturation threshold is initially very high and decreases rapidly. 
This means that at very young ages (younger than maybe 10 to 20~Myr), rotation is not a relevant factor for determining a star's activity. 
For example, the X-ray luminosities of stars in very young clusters depend on mass and age (G\"udel et al. \cite{Guedel07}; Gregory et al. \cite{Gregory16}).
On the main-sequence, the saturation threshold is at approximately 15$\Omega_\odot$ for solar mass stars and is at lower rotation rates for lower mass stars.
This means that low mass stars must reach slower rotation rates before spin-down related activity decay starts, which contributed to lower mass stars remaining highly active longer. 

The dependence of X-ray emission on stellar rotation, mass, and age are well represented as a dependence between the X-ray luminosity normalised to the bolometric luminosity, \mbox{$R_\mathrm{X} = L_\mathrm{X} / L_\mathrm{bol}$}, and the Rossby number, $Ro$, which is a dimensionless parameter defined as \mbox{$Ro = P_\mathrm{rot} / \tau_\mathrm{c}$}, where $P_\mathrm{rot}$ is the rotation period and $\tau_\mathrm{c}$ is the convective turnover time. 
Note that several studies have discussed the possibility that using rotation period instead of Rossby number is better for describing the activity--rotation relation (Reiners et al. \cite{Reiners14}; Magaudda et al. \cite{Magaudda20}).
The $Ro$--$R_\mathrm{X}$ relation is shown in Fig.~\ref{fig:RoRx} and can be described as a broken power-law given by
\begin{equation} \label{eqn:RoRx}
R_\mathrm{X} = \left \{
\begin{array}{ll}
c_1 Ro^{\beta_1}, & \mathrm{if}~~  Ro \ge Ro_\mathrm{sat},\\
c_2 Ro^{\beta_2}, & \mathrm{if}~~  Ro \le Ro_\mathrm{sat},\\
\end{array} \right.
\end{equation}
where $Ro_\mathrm{sat}$ is the saturation Rossby number and $c_1$, $c_2$, $\beta_1$, and $\beta_2$ are constants that can be determined empirically.
If we assume that $R_\mathrm{X}$ is constant in the saturated regime, as is common in the literature, then \mbox{$\beta_1=0$} and \mbox{$c_1 = R_\mathrm{X,sat}$}.
Making this assumption, Wright et al. (\cite{Wright11}) found \mbox{$\beta_2 = -2.18$}, \mbox{$Ro_\mathrm{sat}=0.13$}, and \mbox{$R_\mathrm{X}=10^{-3.13}$}, though depending on the method used for fitting in the unsaturated regime, values for $\beta_2$ between $\sim$2 and 3 are often found (Reiners et al. \cite{Reiners14}).
A weak dependence between $Ro$ and $R_\mathrm{X}$ can be seen in the saturated regime too and there is evidence from extremely rapidly rotators for `supersaturation', which is a decrease in X-ray emission with increasing rotation at the high rotation part of the saturated regime (Jardine et al. \cite{Jardine04}; Argiroffi et al. \cite{Argiroffi16}). 

It is not yet fully clear how well pre-main-sequence and fully convective main-sequence stars follow the same $Ro$--$R_\mathrm{X}$ relation as higher mass main-sequence stars.
In Fig.~\ref{fig:RoRx}, I show the $Ro$--$R_\mathrm{X}$ relation for both pre-main-sequence and main-sequence stars, including for main-sequence stars also fully convective M dwarfs in the unsaturated regime.
For pre-main-sequence stars, the question is difficult to answer because in most very young stellar clusters, most stars are saturated regardless of their rotation rate due to the very large convective turnover times.
Argiroffi et al. (\cite{Argiroffi16}) studied the X-ray--rotation relation in the 12~Myr old cluster h~Per which is useful because the age of this cluster means that many of the slowest rotators could be unsaturated.
Their results suggest a much shallower relation between $Ro$ and $R_\mathrm{X}$ (meaning a smaller $\beta_2$) at this age in the unsaturated regime though this depends on the convective turnover times used and Fig.~\ref{fig:RoRx} suggests that the $Ro$--$R_\mathrm{X}$ relation in h~Per is consistent with the main-sequence relation. 
For fully convective main-sequence M~dwarfs, the difficulty is similarly that the long convective turnover times mean that all but the very slow rotators are in the saturated regime making it hard to characterise the unsaturated regime.
This was studied for the young open cluster NGC~2547 by Jeffries et al. (\cite{Jeffries11}) who found that the unsaturated regimes for fully convective M~dwarfs and higher mass stars are similar as can be seen in Fig.~\ref{fig:RoRx}, and several studies have backed up their conclusions (Wright et al. \cite{Wright16}; Wright et al. \cite{Wright18}).
There is evidence that the slope for fully convective stars is steeper (Magaudda et al. \cite{Magaudda20}) which might not be visible in Fig.~\ref{fig:RoRx} since I do not include upper limits.

\begin{figure}
\centering
\includegraphics[width=0.49\textwidth]{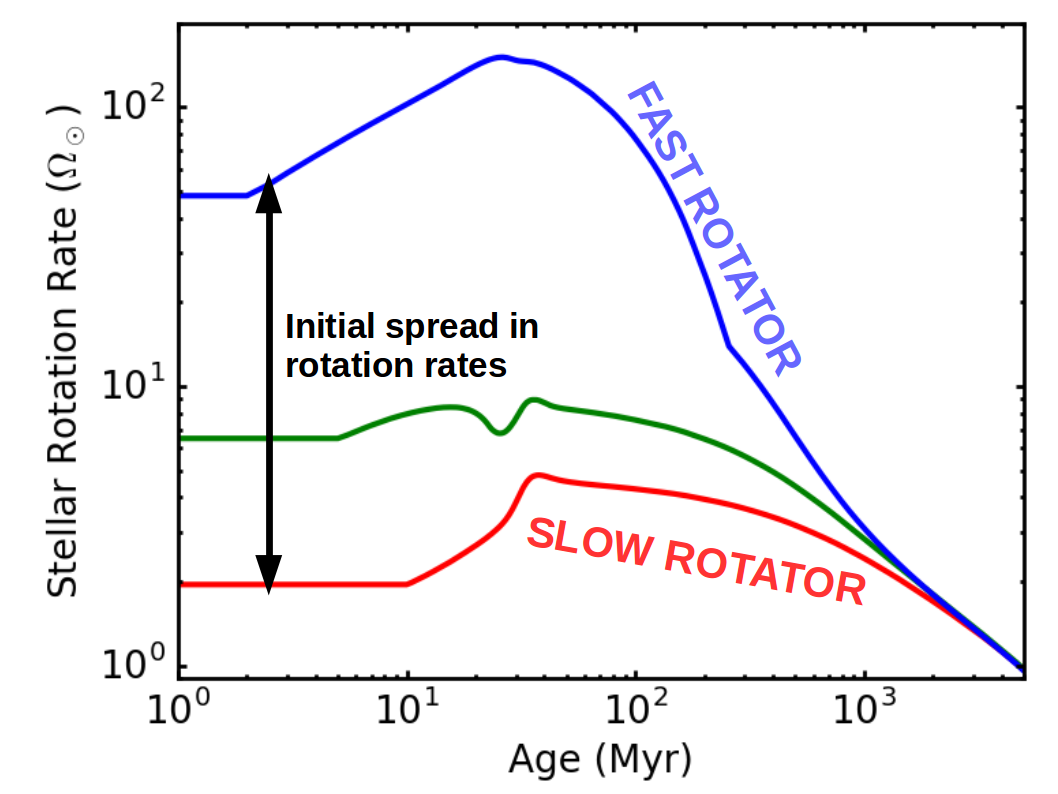}
\includegraphics[width=0.49\textwidth]{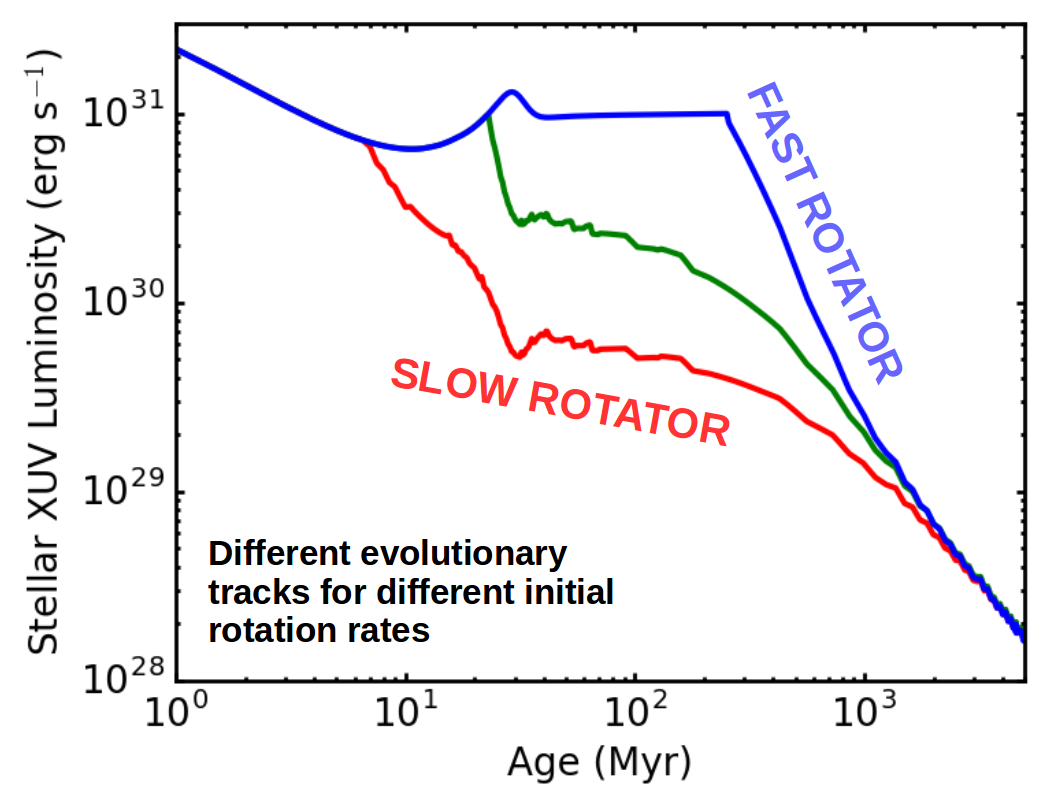}
\vskip -0mm
\caption{
\emph{
Evolutionary tracks for rotation (left) and XUV emission (right) from Tu et al. (\cite{Tu15}) for solar mass stars with different initial rotation rates.
Here, XUV is calculated considering just X-ray and EUV wavelengths.
}
}
\label{fig:rotxuvattracks}
\end{figure}

\subsection{Activity evolution}

Given the close link between rotation and activity, we should expect that the activity levels of stars decay with age due to rotational spin-down.
This was observed by Skumanich (\cite{Skumanich72}) who found that Ca II emission decays simultaneously with rotation over billions of years.
For X-ray emission from solar mass stars, this decay was characterised by G\"udel et al. (\cite{Guedel97}) who found \mbox{$L_\mathrm{X} = 2.1 \times 10^{28} t^{-1.5}$~erg~s$^{-1}$}, where $t$ is the age in Gyr.
This decay in high energy emission is seen for all magnetic activity related emission, including extreme ultraviolet, though at longer wavelengths the decay law is less steep (Ribas et al. \cite{Ribas05}; Claire et al. \cite{Claire12}). 
This is in part due to the fact that more active stars have hotter emitting plasma (Schmitt et al. \cite{Schmitt97}; Johnstone \& G\"udel \cite{JohnstoneGuedel15}), meaning that the spectrum is more shifted to shorter wavelengths.
If we consider again the link between rotation and high-energy emission and the various possible rotational evolution tracks that stars can follow (Fig.~\ref{fig:rottracks}), we can see that there is a problem with describing activity decay as a single unique power-law. 
Since stars born as fast rotators remain rapidly rotating for much longer than those born as slow rotators, we should expect that the initial rotation rate of a star plays an important role in determining the evolution of its emission.
This was shown in Tu et al. (\cite{Tu15}) for solar mass stars, and three different possible evolutionary tracks for X-ray and EUV emission from this study are demonstrated in Fig.~\ref{fig:rotxuvattracks}.
The slow, medium, and fast rotator tracks correspond to the 10$^\mathrm{th}$, 50$^\mathrm{th}$, and 90$^\mathrm{th}$ percentiles of the rotation distribution and the evolution of X-ray emission (0.1--10~nm) for these three tracks can be approximated as
\begin{equation}
\hbox{$
L_{\rm X} = \left\{ 
  \begin{array}{l}
     2.0\times 10^{31} t^{-1.12} \\
     2.6\times 10^{32} t^{-1.42} \\
     2.3\times 10^{36} t^{-2.50} \\
  \end{array}
  \right.
$}\nonumber
\end{equation}
where $L_\mathrm{X}$ is in erg~s$^{-1}$ and $t$ is in Myr.
The evolution of the EUV emission (10--91~nm) for these three tracks can be approximated as
\begin{equation}
\hbox{$
L_{\rm EUV} = \left\{ 
  \begin{array}{l}
     7.4\times 10^{31} t^{-0.96} \\
     4.8\times 10^{32} t^{-1.22} \\
     1.2\times 10^{36} t^{-2.15} \\
  \end{array}
  \right.
$}\nonumber
\end{equation}
where the units of the quantities are the same as the previous equations.

\section{Properties and Evolution of Stellar Winds}

% Basic intro into winds
In the previous section, I concentrate on the evolution of stellar rotation and its connection to magnetic activity without discussing in any detail winds, which is a primary topic of this review.
In this section, I discuss how stellar wind bulk properties -- primarily mass flux and outflow speed -- evolve with time and how this depends on the parameters of the star, such as mass and initial rotation rate.
Unfortunately little is known about this for low-mass stars other than the Sun and what is known is not known with any certainty.

\subsection{Observational constraints on winds}

While observations of the long term evolution of stellar rotation make it clear that winds are ubiquitous among low-mass stars, other than for the case of the solar wind, we do not possess clear observational constraints on their properties. 
Much work has been dedicated to detecting and characterising winds but since winds consist of very tenuous gas, this is a very difficult task.
Most methods that have been developed attempt to indirectly determine wind properties by measuring the effects of these winds on the local interstellar environment around the star, planetary or stellar companions, or the star itself.

\subsubsection{Rotational evolution} \label{sect:windobsrot}

\begin{figure}
\centering
\includegraphics[width=0.7\textwidth]{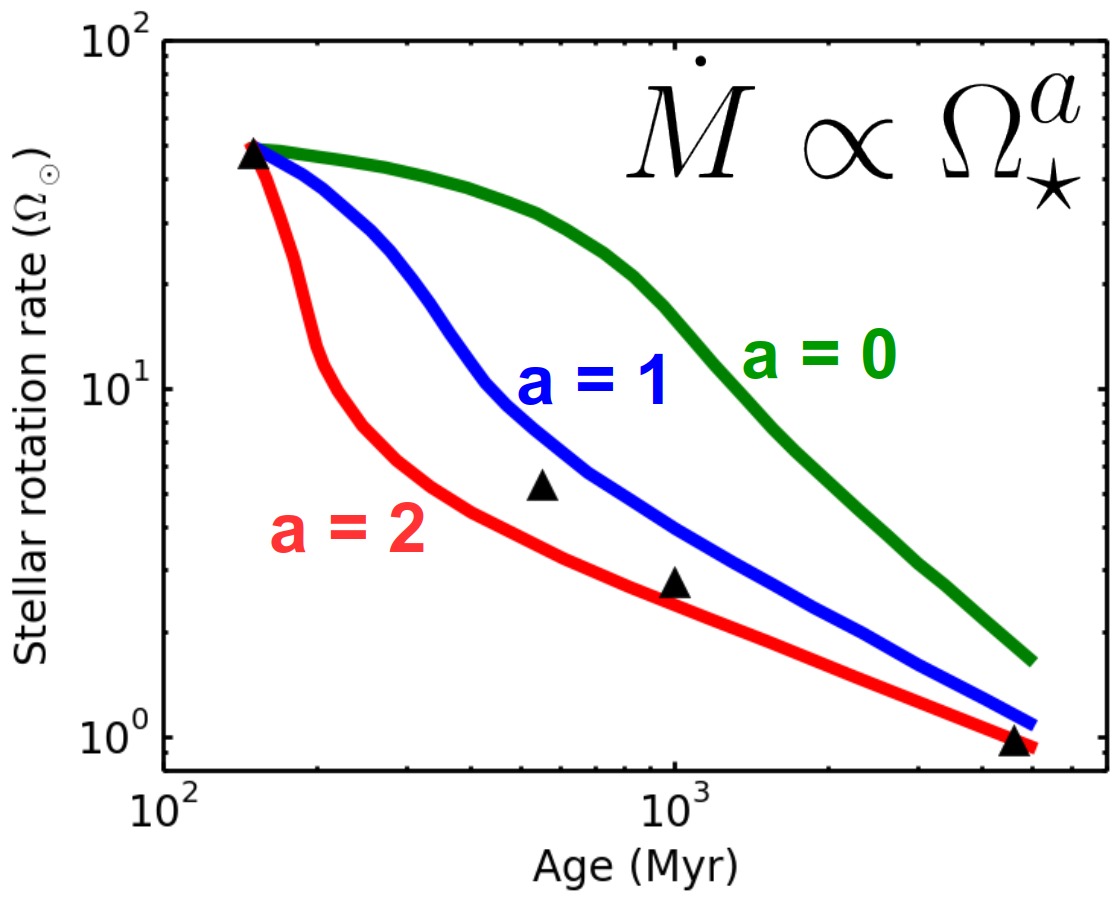}
\vskip -4mm
\caption{
\emph{
Figure demonstrating how the observed evolution of stellar rotation can be used to constrain the properties of stellar winds based on the rotational evolution model of Johnstone et al. (\cite{Johnstone15b}).
The black triangles from show observational constraints on the evolution of stars at the 90$^\mathrm{th}$ percentile of rotation distributions for solar mass stars.
The rotation tracks start at the first constraint and show cases for different values of $a$ in the assumed equation $\dot{M}_\mathrm{w} \propto \Omega_\star^a$.
The best fit value of $a$ was found to be 1.33 for this particular rotational evolution model.
}
}
\label{fig:rotfitting}
\end{figure}

The rotational evolution of stars is the most clear indication that stellar winds are present on low-mass stars and possibly has the most potential for allowing us to learn about wind properties.
For example, the observed differences between the spin-down rates of saturated and unsaturated stars discussed in Section~\ref{sect:rotepochs} is a good indication that the mass loss rates of winds saturate at rapid rotation with other activity related phenomena such as X-ray emission.  
The rate at which a star spins down on the main-sequence depends on several factors, including its mass and radius, the properties of its magnetic field, and the properties of its wind and it is therefore possible to determine wind properties from the observed rotational evolution of stars as explored by several studies (Gaidos et al. \cite{Gaidos00}; Matt et al. \cite{MattPudritz08a}).
The model of Weber \& Davis (\cite{WeberDavis67}) for angular momentum loss from the winds of magnetised rotating stars gives the angular momentum loss from a stellar wind as \mbox{$\tau_\mathrm{w} \propto \dot{M}_\mathrm{w} \Omega_\star R_\mathrm{A}^2$}, where $\dot{M}_\mathrm{w}$ is the wind mass loss rate, $\Omega_\star$ is the stars rotation rate, and $R_\mathrm{A}$ is the Alfv\'en radius. 
Past ages of approximately 1~Gyr, the spin-down of solar mass stars can be approximated as \mbox{$\Omega_\star \propto t^{-0.5}$} (Skumanich \cite{Skumanich72}), where $t$ is the age, implying that \mbox{$\dot{\Omega}_\star \propto \Omega_\star^{3}$}.
If we assume approximately constant values for the stellar moment of inertia and the Alfv\'en radius, this implies that \mbox{$\dot{M}_\mathrm{w} \propto \Omega_\star^2 \propto t^{-1}$}.
While this reasoning is overly simplistic, it demonstrates the potential to derive wind properties from observed rotational evolution. 

Ideally, we would determine wind properties using a full detailed rotational evolution model fit to the abundant observational constraints on rotation distributions in young stellar clusters.
Using such a model, Matt et al. (\cite{Matt15}) found that \mbox{$\dot{M}_\mathrm{w} \propto M_\star^{1.3} Ro^{-2}$}, where $Ro$ is the Rossby number. 
Similarly, considering only stars after the first 100~Myr, Johnstone et al. (\cite{Johnstone15b}) found \mbox{$\dot{M}_\mathrm{w} \propto R_\star^{2} \Omega_\star^{1.33} M_\star^{-3.36}$}.
A plausible assumption is that the surface mass flux has power-law dependences on Rossby number and mass, which gives
\begin{equation} \label{eqn:MdotRossby}
\dot{M}_\mathrm{w} \propto R_\star^2 Ro^a M_\star^b,
\end{equation}
where $a$ and $b$ are parameters to be determined.
Note that the above is only when $Ro$ is greater than the saturation Rossby number, $Ro_\mathrm{sat}$, and when this is not the case, $Ro_\mathrm{sat}$ should be used in place of $Ro$.
Johnstone et al. (submitted) used measurements of stellar rotation and activity to constrain these parameters and found \mbox{$a=-1.76$}, \mbox{$b=0.649$}, and \mbox{$Ro_\mathrm{sat}=0.0605$}.
This equation can be used to find the mass loss rate of any star in the mass range 0.1 to 1.2~M$_\odot$ by using the solar mass loss rate of \mbox{$\dot{M}_\odot = 1.4 \times 10^{-14}$~M$_\odot$~yr$^{-1}$} and the convective turnover times derived by Spada et al. (\cite{Spada13}), though there is much uncertainty for the determination of these parameters.
This suggests that \mbox{$\dot{M}_\mathrm{w} \propto t^{-0.88}$} for the solar wind after the first Gyr.
The estimates for $\dot{M}_\mathrm{w}$ given here are possibly the best observationally driven constraints on wind mass loss rates available, but they still ignore many of the details of rotational spin-down physics, such as changes in the magnetic field properties and wind acceleration.

\subsubsection{Radio observations}

Several attempts have been made to observe stellar winds at radio wavelengths, and while such winds have not been directly detected, this technique has been able to put important upper limits on mass loss rates.
These upper limits can be derived in two ways.
Given their temperatures and ionization states, it is expected that free-free thermal radiation at radio wavelengths will be detectable if the winds are sufficiently luminous (Panagia \& Felli \cite{PanagiaFelli75}; Wright \& Barlow \cite{WrightBarlow75}).
The non-detection of radio emission from a star's wind can therefore be used to derive an upper limit on the wind's density and mass loss rate.
For example, Gaidos et al. (\cite{Gaidos00}) was unable to detect emission from the winds of three young Sun-like stars and found that the mass loss rates must be below $\sim 5 \times 10^{-11}$~M$_\odot$~yr$^{-1}$.
Possibly even more useful is the fact that with a sufficiently dense wind, radiation emitted from a star's surface at some radio wavelengths will not escape the system. 
The observations of flares on the surfaces of stars in these wavelength regions imply that the winds are not sufficiently dense, which can be used to derive upper limits on mass loss.
This method was used by Lim \& White (\cite{LimWhite96}) to derive an upper limit on the mass loss of the 0.3~M$_\odot$ M-dwarf YZ~CMi of $\sim 10^{-12}$~M$_\odot$~yr$^{-1}$.
Note that in some cases, stellar winds can also cause radio emission from the magnetospheres of planets with short orbital periods to be unobservable (Vidotto \& Donati \cite{VidottoDonati17}).
The most sensitive upper limits on the winds of young solar mass stars were derived by Fichtinger et al. (\cite{Fichtinger17}) who found that the mass loss rates of the 300~Myr old $\pi^1$~UMa and the 650~Myr old $\kappa^1$~Ceti are below $\sim5 \times 10^{-12}$~M$_\odot$~yr$^{-1}$.
These upper limits suggest that very active stars likely have mass loss rates that are within a factor of 100 of that of the current Sun, though they say little about the winds of very young pre-main-sequence stars.

\subsubsection{Interactions with the interstellar medium}

It is possible that wind properties can be inferred from their interactions with the interstellar medium (ISM) and stellar astrospheres, which are the stellar equivalents of the heliosphere.
At distances of several tens of AU from the Sun, the solar wind begins to interact strongly with neutral hydrogen from the ISM. 
Protons in the solar wind charge exchange with ISM hydrogen creating a slowly moving ion and a neutral hydrogen atom traveling at several hundred km~s$^{-1}$ called an energetic neutral atom (ENA).
This process leads to the emission of an X-ray photon and Wargelin \& Drake (\cite{WargelinDrake02}) attempted to observe these photons in the astrosphere of Proxima Centauri. 
Their lack of detection implied an upper limit on the mass loss rate of $\sim 3 \times 10^{-13}$~M$_\odot$~yr$^{-1}$.

The slowly moving ion created in charge exchange reactions is picked up by the solar wind and this has the effect of slowing the wind and eventually making it subsonic.
If enough neutral hydrogen is present in the region of the ISM that a stellar system is embedded in, a wall of neutral hydrogen can build up at the edge of the astrosphere (Wood et al. \cite{Wood01}).
If our viewing angle to the star is such that the star's radiation passes through this neutral hydrogen wall, much of the star's Ly-$\alpha$ emission line will be absorbed.
Unfortunately, due to absorption by neutral hydrogen in the ISM, only a small fraction of a star's Ly-$\alpha$ emission is observable, which makes detecting the signature of this astrospheric absorption challenging.
The results can be compared to models of wind--ISM interactions and used to estimate wind properties. 
Wood et al. (\cite{Wood01}, \cite{Wood02}, \cite{Wood05}, \cite{Wood14}) used this technique to estimate wind properties of several stars and found that stars with stronger X-ray emission have higher mass loss rates.
For example, they estimated a mass loss rate for the 0.82~M$_\odot$ K~dwarf $\epsilon$~Eri to be $\sim 6 \times 10^{-13}$~M$_\odot$~yr$^{-1}$, which is a factor of 30 above that of the current Sun.
They results suggested that \mbox{$\dot{M}_\mathrm{w} \propto F_\mathrm{X}^{1.34}$}, which for the solar wind at later ages suggests \mbox{$\dot{M}_\mathrm{w} \propto t^{-2.33}$}. 

The derived relation between X-ray activity and mass loss breaks down for the most active stars which all show much weaker than expected mass loss.
For example, their measured mass loss from the young 300~Myr old solar analogue $\pi^1$~UMa was half that of the modern solar wind (Wood et al. \cite{Wood14}).
These results suggest that very active stars could be have much weaker winds than we would expect, and could mean that for the first Gyr of the solar system's history, the solar wind mass flux was similar to modern values.
However, as discussed in Johnstone et al. (\cite{Johnstone15b}), this interpretation is difficult to reconcile with our empirical understanding of stellar rotational evolution. 
The rapid spin-down of rapidly rotating active stars that we infer from young stellar clusters requires high wind mass loss rates. 
The dependence of the solar wind mass loss rate on $t^{-2.33}$ at later ages is also much stronger than the results implied by rotational evolution models which suggest a $t^{-1}$ dependence.

\subsubsection{Effects on planetary companions} \label{sect:windobsplanet}

%\fbox{\includegraphics[trim = 55mm 25mm 55mm 15mm, clip=true,width=0.29\textwidth]{Plots/Cloud2_xy.pdf}}

\begin{figure}
\centering
\fbox{\includegraphics[trim = 0mm 0mm 0mm 0mm, clip=true,width=0.5\textwidth]{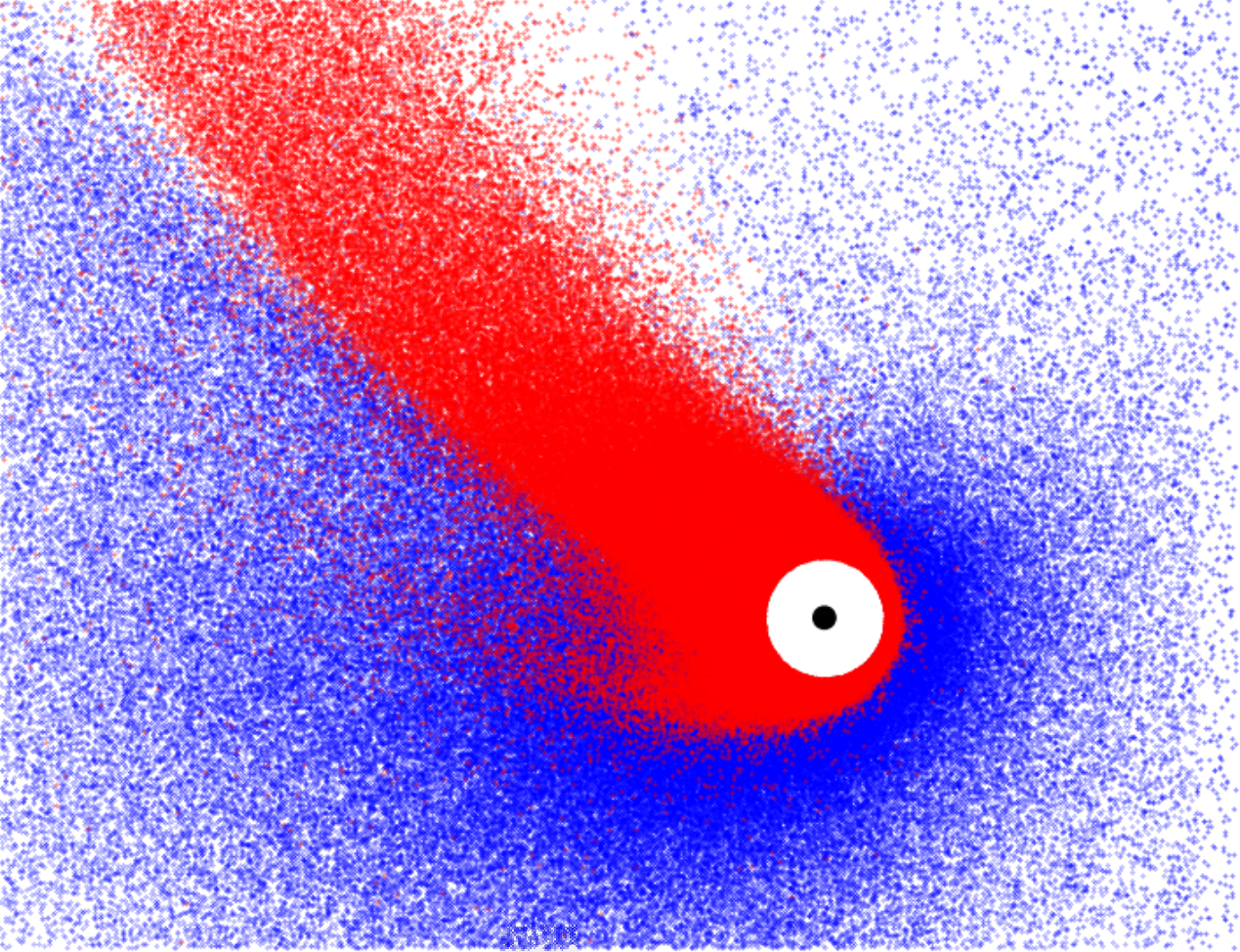}}
\includegraphics[trim = 0mm 0mm 0mm 0mm, clip=true,width=0.85\textwidth]{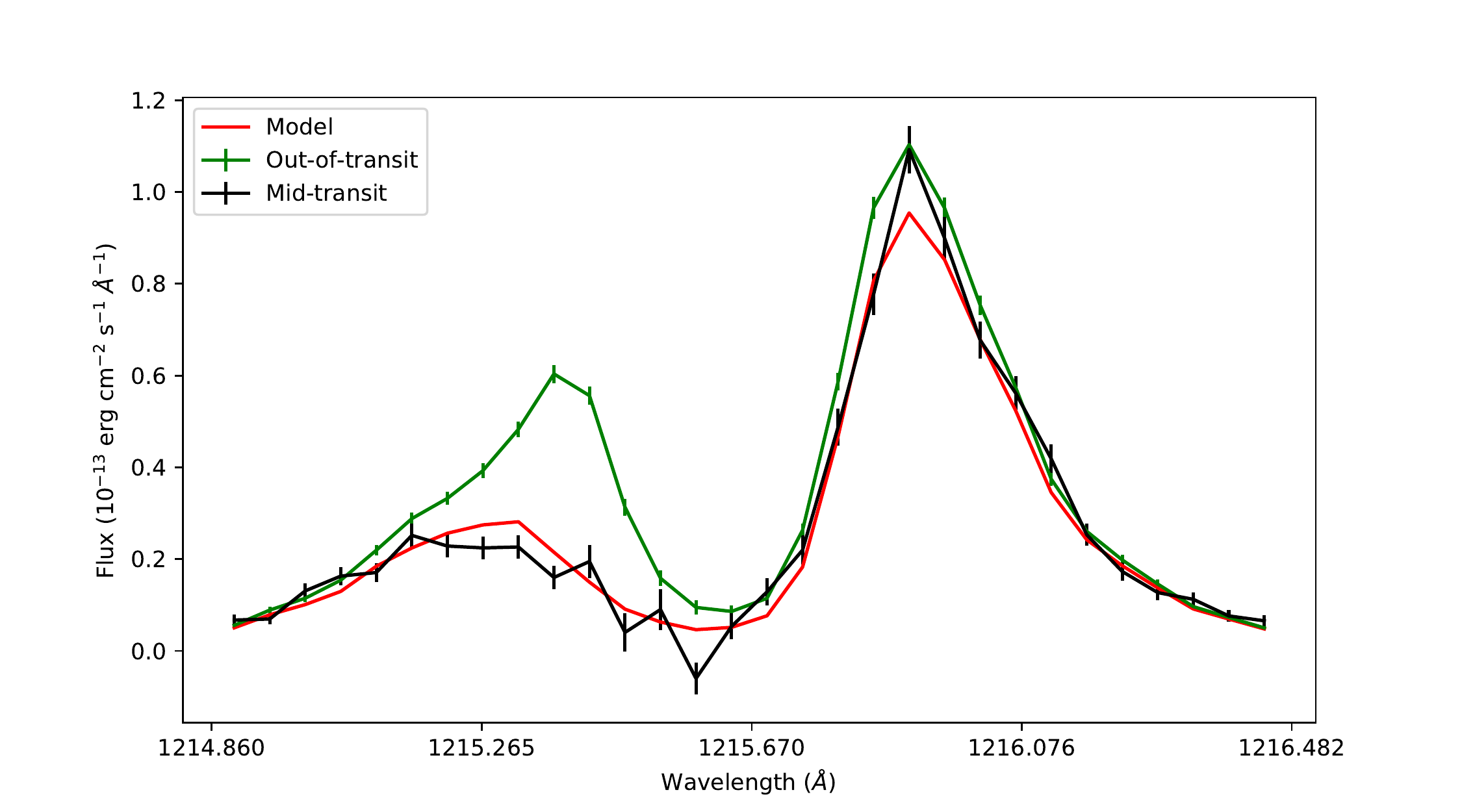}
\vspace{-5mm}
\caption{
\emph{
The atomic hydrogen cloud around the planet Gliese~436b, with red and blue showing neutral and ionised hydrogen atoms (each dot corresponds to \mbox{$1 \times 10^{32}$ atoms}) and the black circle showing the planet's size.  
The lower panel shows the observed Ly-$\alpha$ emission line of the star in and out of transit and the mid-transit model simulated by Kislyakova et al. (\cite{Kislyakova19}).
Images courtesy of K. G. Kislyakova.
}
}
\vspace{-5mm}
\label{fig:lyatransit}
\end{figure}

Several techniques have been suggested for inferring wind properties indirectly using measurements of the influences of these winds on transiting planets.
Planets are surrounded by a large and tenuous cloud of particles originating from their upper atmospheres called the exosphere and neutral particles in the exosphere charge exchange with stellar wind protons creating an tail of ENAs flowing away from the planet, as shown in Fig.~\ref{fig:lyatransit}.
While charge exchange produces X-ray photons, this emission is unobservable outside our solar system (Kislyakova et al. \cite{Kislyakova15}).
Neutral hydrogen atoms in both the exosphere and in the ENA tail can transit the star and the resulting absorption causing variability of the star's Ly-$\alpha$ emission line over the course of each orbit.
This can be seen for example in the Ly-$\alpha$ line of the M~dwarf Gliese~436 which is orbited by a warm Neptune-sized transiting planet as shown in Fig.~\ref{fig:lyatransit}.
This absorption has been observed for several transiting planets with hydrogen dominated primordial atmospheres (Vidal-Madjar et al. \cite{VidalMadjar03}; Ehrenreich et al. \cite{Ehrenreich08}, \cite{Ehrenreich15}).

The time variability of the Ly-$\alpha$ line for a given star--planet system depends on the properties of the star's wind in the vicinity of the planet and it is possible to use this to derive knowledge about the winds of such stars.
Kislyakova et al. (\cite{Kislyakova14b}) compared the Ly-$\alpha$ line of HD~209458, which is transited by a short period planet with a mass slightly below that of Jupiter, to models for interactions between the star's wind and the planet's exosphere.
They were able to estimate a wind speed in the region of the planet of 400~km~s$^{-1}$ and they were also able to estimate the magnetic moment of the planet which is important because measurements of exoplanet magnetic moments are very challenging (Weber et al. \cite{Weber17}).
Similarly, Bourrier et al. (\cite{Bourrier16}) derived a stellar wind speed in the vicinity of the transiting planet GJ~436b of 85~km~s$^{-1}$ and Vidotto \& Bourrier (\cite{VidottoBourrier17}) used their constraints to model the host star's wind, finding a mass loss rate of \mbox{$\sim 10^{-15}$~M$_\odot$~yr$^{-1}$}.
Other methods for constraining stellar wind properties using transiting exoplanets have been explored (Llama et al. \cite{Llama13}).

Another promising avenue for measuring stellar wind properties involves hydrogen atmosphere white dwarfs that have spectral absorption lines from metals, which in many cases is unexpected given that metals should rapidly diffuse out of the upper atmosphere. 
It is possible that in many systems, these metals could come from the winds of companion M~dwarfs accreted onto the surfaces of the white dwarfs (Parsons et al. \cite{Parsons12}).
Debes (\cite{Debes06}) explored the possibility of constraining the mass loss of the M~dwarf companions of six such white dwarfs, three of which have small orbital separations of 0.015~AU or less and the other three have separations of 1.5--159~AU.
For the close orbit systems, they estimated mass loss rates between $10^{-16}$ and \mbox{$6 \times 10^{-15}$~M$_\odot$~yr$^{-1}$} and for the wider systems, they estimated values above \mbox{$10^{-10}$~M$_\odot$~yr$^{-1}$}, which they argue are unrealistically large.

\begin{figure}[!hp]
\centering
\includegraphics[width=0.8\textwidth]{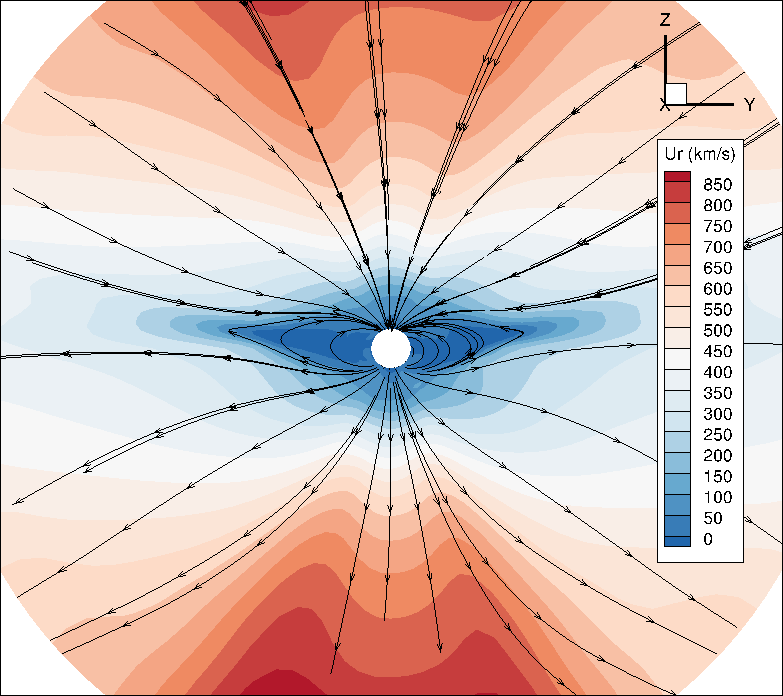}
\hspace*{1.0mm}\includegraphics[width=0.8\textwidth]{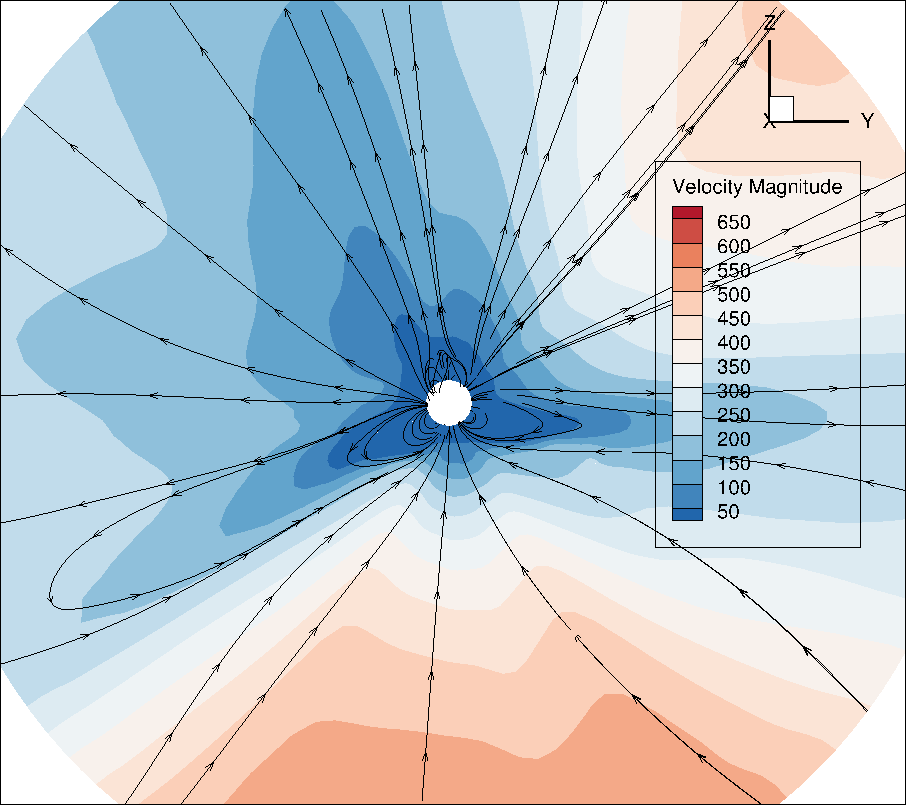}
\vspace{-2mm}
\caption{
\emph{
Magnetic field and velocity structure in the solar wind with dipolar (upper panel) and more complex (lower panel) field structures as simulated by Boro-Saikia et al. (\cite{BoroSaikia20}) using the Space Weather Modelling Framework (T\'oth et al. \cite{Toth05}).
Black lines show magnetic field lines and colored contours show wind speed and the close relation between acceleration and magnetic field can be seen.
Images courtesy of S. Boro-Saikia.
}
}
\vspace{-4mm}
\label{fig:solarwind}
\end{figure}

\subsection{Solar wind: basic properties}

Although a simplification, it is useful here to describe the solar wind as being composed of three components: these are the slow wind, the fast wind, and coronal mass ejections (CMEs).
Typically, the solar wind is not isotropic but has variable speed in different directions with speeds typically between 300 and 900~km~s$^{-1}$ and the distribution of wind speeds shows two clear components with average speeds of 400~km~s$^{-1}$ for the slow component and 760~km~s$^{-1}$ for the fast component.
The slow wind is cooler and more variable on short timescales than the fast wind and the two components have different abundances of heavy ions (von Steiger et al. \cite{vonSteiger10}).
However, the two components have very similar mass fluxes, meaning that the outflow mass flux from the Sun is approximately isotropic and the mass loss rate is approximately constant over the solar cycle.

\begin{figure}
\centering
\hspace*{1.0mm}\includegraphics[width=0.99\textwidth]{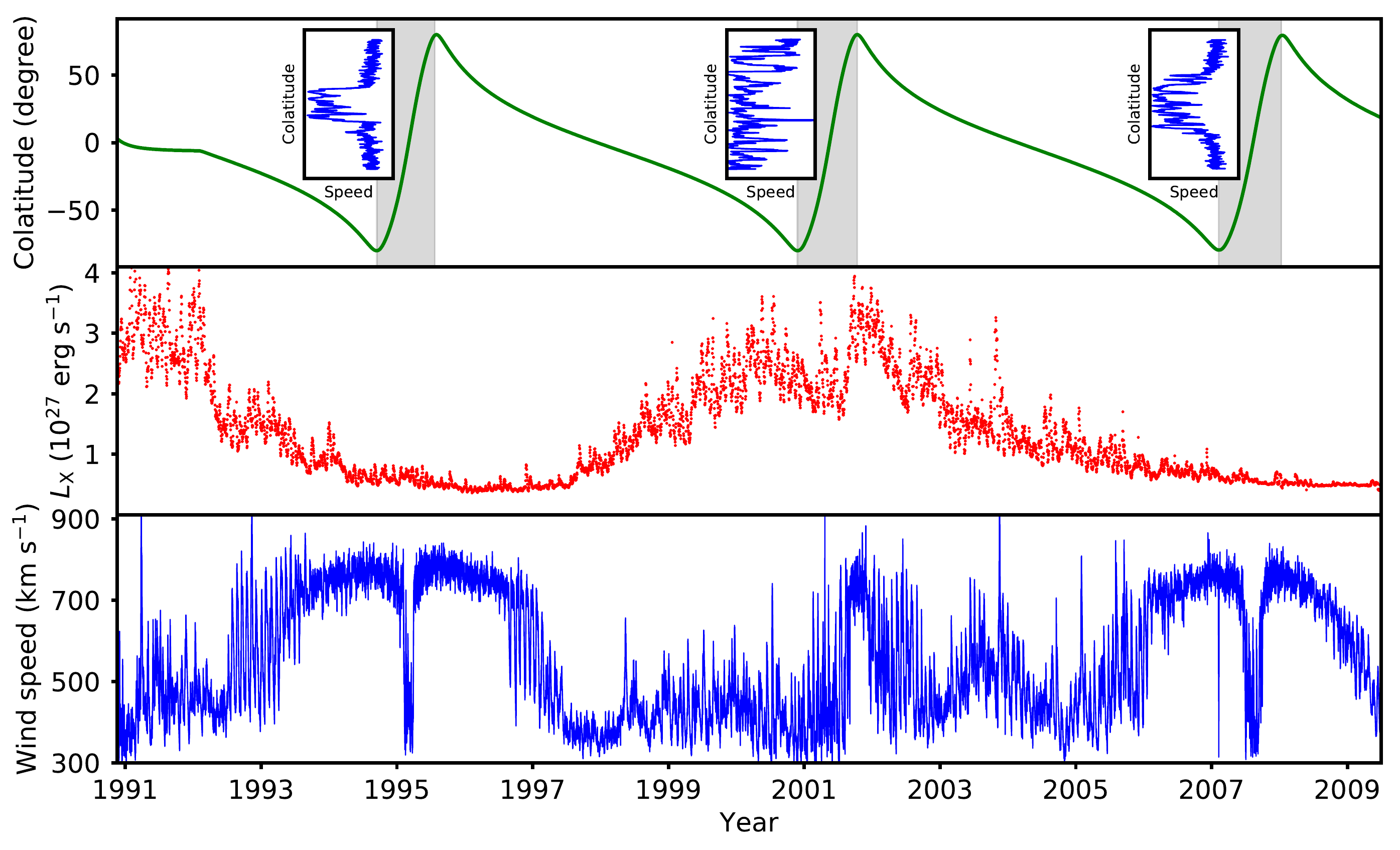}
\vspace{-8mm}
\caption{
\emph{
Solar wind speed as a function of latitude and solar cycle as revealed by the Ulysses spacecraft between 1990 and 2010, with Ulysses measurements obtained from \href{cohoweb.gsfc.nasa.gov}{http://cohoweb.gsfc.nasa.gov}.. 
The upper panel shows the colatitude of Ulysses and the shaded areas indicate the three fast latitude scans.
The insets next to each fast latitude scan shows the wind speed measured by the spacecraft as a function of colatitude for the three scans and we can easily see the differences in wind structure between cycle minimum (first and third scans) when the Sun's global magnetic field was dipolar and maximum (second scan) when the field was more complex.
The middle panel shows solar X-ray luminosity during this time, with the $L_\mathrm{X}$ values derived from the Flare Irradiance Spectrum Model (Chamberlin et al. \cite{Chamberlin07}) and the solar cycle variability is clearly visible.
The lower panel shows wind speed measured by Ulysses during the time.
}
}
\vspace{-4mm}
\label{fig:ulysses}
\end{figure}

If we do not consider transitory events like CMEs (discussed instead in Section~\ref{sect:cmes}), the main change in the solar wind properties over the solar cycle is the spatial distribution of slow and fast wind.
These two components are closely related to the geometry of the Sun's global magnetic field, which is typically very simple and dipolar at activity minimum and much more complex and multipolar at activity maximum.
As can be seen in Fig.~\ref{fig:solarwind}, we can break down the Sun's global magnetic field into regions of closed and open field lines and it is known empirically that fast wind originates from regions of open mangetic field, whereas the slow wind originates from above closed magnetic field (Wang \& Sheeley \cite{Wang90}).
However, despite this clear correlation, the exact physical origin on the Sun's surface of the slow component is unclear and various possibilities, such as the boundaries between open and closed field lines and active regions within closed field regions, have been suggested (Abbo et al. \cite{Abbo16}).
At solar minimum, fast wind originates from high latitudes and slow wind dominates in equatorial regions, whereas at solar maximum, the distribution of fast and slow wind is much more complex and unpredictable. 
This can be seen from the wind speed measurements of the Ulysses spacecraft which which had a very high orbital inclination relative to the plane of the solar system and was therefore able to measure the solar wind properties at all latitudes.
Wind speed measurements and the relation to latitude and the solar cycle from \emph{Ulysses} are shown in Fig.~\ref{fig:ulysses}.

\subsection{Wind physics}

\begin{figure}
\centering
\hspace*{1.0mm}\includegraphics[width=0.9\textwidth]{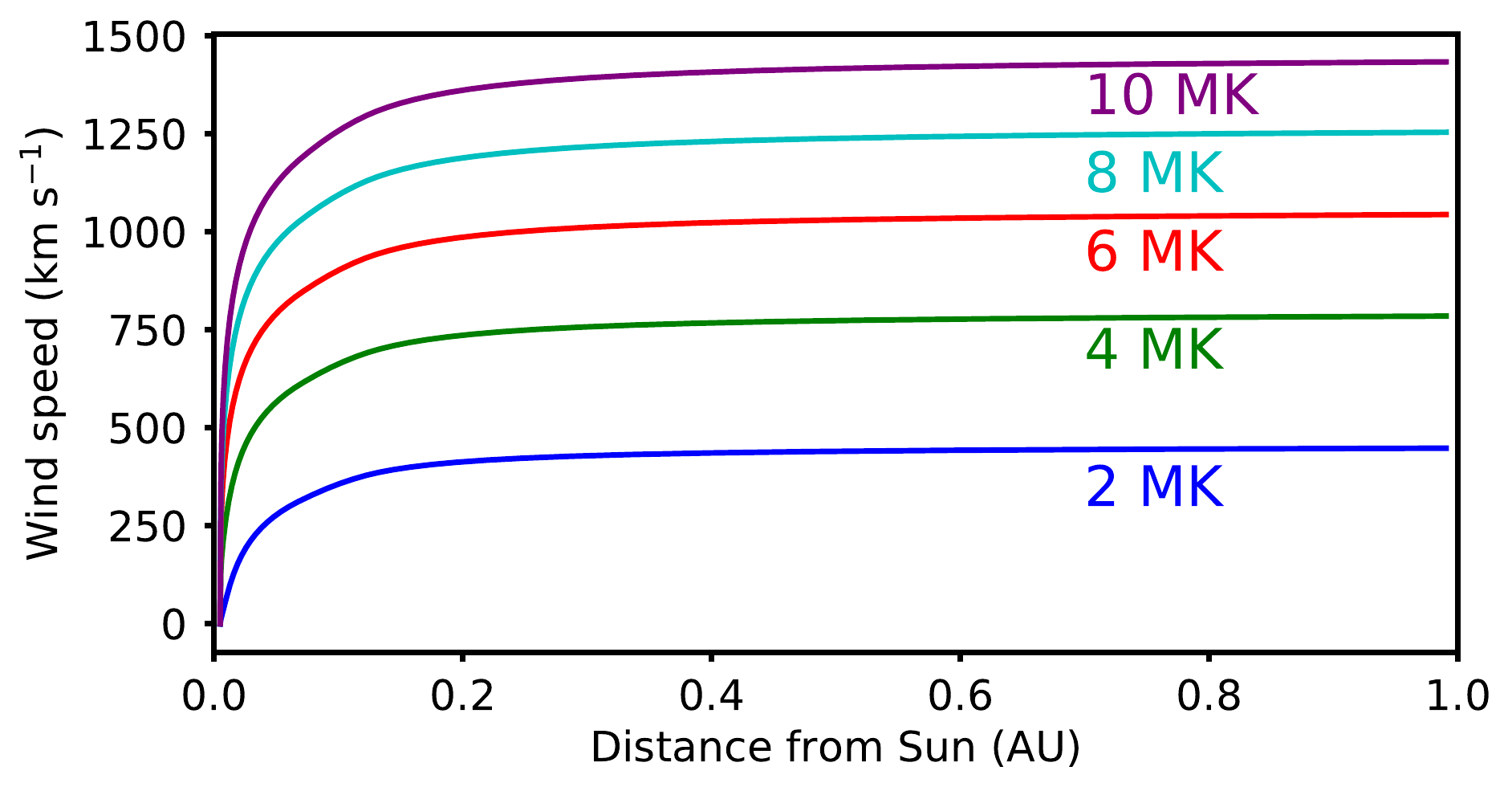}
\vspace{-3mm}
\caption{
\emph{
Wind speed as a function of distance from the center of the Sun for different lower boundary temperatures assuming a thermal pressure driven wind only and a polytropic equation of state as calculated using the model of Johnstone et al. (\cite{Johnstone15a}).
}
}
\vspace{-4mm}
\label{fig:windveltgas}
\end{figure}

% Basic Parker and WD model
After the initial suggestion by Biermann (\cite{Biermann52}) of the presence of a fast flowing wind propagating outwards from the Sun, Parker et al. (\cite{Parker58}) derived a simple hydrodynamic model for the radial outflow of the solar wind. 
In this model, the wind is accelerate away from the Sun by thermal pressure gradients, which is possible given that the outer atmosphere of the Sun reaches temperatures of million of K and the assumption by Parker et al. (\cite{Parker58}) that the wind is isothermal.
This model was extended to take into account the effects of the Sun's rotation and magnetic field on the wind by Weber \& Davis (\cite{WeberDavis67}) who derived for the radial acceleration of the wind
\begin{equation} \label{eqn:wdmodel}
V_r \frac{d V_r}{dr} = - \frac{1}{\rho} \frac{dp}{dr} - \frac{G M_\star}{r^2} + \frac{V_\phi^2}{r} - \frac{B_\phi}{4\pi \rho r} \frac{d}{dr} \left( r B_\phi \right),
\end{equation}
where $r$ is the distance from the center of the star, $V_r$ is the radial outflow speed,  $\rho$ is the mass density, $p$ is the thermal pressure, and $V_\phi$ and $B_\phi$ are the azimuthal components of the wind velocity and magnetic field.
The four terms on the right hand side of this equation correspond to the acceleration due to thermal pressure gradients, gravity, the centrifugal force, and the radial component of the Lorentz force.
Due to the Sun's slow rotation and relatively weak magnetic field, the third and forth terms are negligible for the solar wind and when ignored, this equation gives the radial acceleration from the model by Parker et al. (\cite{Parker58}).
In this model, the wind speed far from the surface of the star is determined primarily by temperature, with higher temperature winds having higher outflow speeds.
The mass flux in the wind is determined by both the temperature and the density at the base, wherever that is defined.
The isothermal assumption leads to unrealistically large acceleration far from the Sun and a slightly more realistic model can be obtained by assuming a polytropic equation of state, meaning that \mbox{$p \propto \rho^\alpha$}, where \mbox{$\alpha = 1$} and \mbox{$\alpha = 5/3$} correspond to isothermal and adiabatic winds respectively (though a Parker wind cannot form if the gas is adiabatic). 
Note that this is not the relation \mbox{$p \propto \rho^\gamma$} for an adiabatic gas and $\alpha$ is not the adiabatic index.
Within the solar corona, $\alpha$ is approximately 1.05 (Suess et al. \cite{Suess77}) and in the inner heliosphere, $\alpha$ is approximately 1.46 (Totten et al. \cite{Totten95}).
Wind acceleration profiles for a polytropic wind with different base temperatures are shown in Fig.~\ref{fig:windveltgas} and a detailed explanation of these types of models is given by Lamers \& Cassinelli (\cite{LamerCassinelli99}).

% Heating mechanisms
The main problem with the models described above is the lack of any description of how the wind is heated, with the temperature and density at the base of the wind being free parameters. 
Since we have no clearly reliable method for determining these parameters for other stars, a primary goal for solar and stellar wind physics is the identification and description of the mechanisms responsible for heating the outer atmospheres and winds of stars. 
It is generally accepted that the source of the energy is convective motions in the photosphere and this energy is transferred to and released in the corona and wind by magnetic fields, but the mechanisms responsible are currently disputed.
As described in detail in Cranmer (\cite{Cranmer09b}), physical wind heating models break down into two classes: these are wave/turbulence driven models and reconnection/loop-opening models. 
The former involve waves created by the jostling of magnetic field lines in the photosphere propagating upwards and being dissipated at higher altitudes. 
These are typically assumed to be Alv\'en waves and many physical models that heat winds in this way have been developed (Cranmer \cite{Cranmer07}; Suzuki et al. \cite{Suzuki13}; van Ballegooijen \& Asgari-Targhi \cite{vanBallegooijen16}).
The latter involve the dragging around of magnetic field lines by photospheric convective motions causing the field lines to become stressed, building up energy in the magnetic fields that is then released by reconnection. 
Currently Alfv\'en wave heating is the favored explanation and is used as the main wind driving mechanism in 3D global wind models (van der Holst et al. \cite{vanderHolst14}; R\'eville et al. \cite{Reville20}).
For a recent review, see Cranmer \& Winebarger (\cite{Cranmer19}).

% Acceleration from waves
A further complication to the picture involves measurements of the solar wind temperatures near and far from the Sun.
Temperatures derived from the Solar Heliospheric Observatory (SOHO) mission have shown that the gas within coronal holes is typically between 1 and 2~MK (Esser et al. \cite{Esser99}; Antonucci et al. \cite{Antonucci00}) and if we assume an isothermal Parker wind, these temperatures are high enough to accelerate the wind to speeds typical for the fast solar wind.
However, \emph{in situ} measurements of the wind far from the Sun show that it is not isothermal and the hotter fast component typically has temperatures of \mbox{$\sim 3 \times 10^5$~K} at 1~AU.
While this is much hotter than we would expect if the solar wind was cooling adiabatically as it expands, meaning that wind heating must take place throughout the inner heliosphere, models that correctly reproduce the temperature structure of the wind are unable to explain the outflow speeds of the fast wind using thermal pressure gradients as the only acceleration mechanism (e.g. see Fig.~4 of Johnstone et al. \cite{Johnstone15a}).
It is therefore necessary that another acceleration mechanism not included in Eqn.~\ref{eqn:wdmodel} is important, and a prime candidate for this is momentum deposition from waves.
Cranmer (\cite{Cranmer04}) estimated that close to the solar surface, thermal pressure gradients dominate the acceleration, and by a few solar radii, direct momentum dissipation from waves contribute approximately half of the acceleration (see their Fig.~3).

\begin{figure}
\centering
\includegraphics[width=0.8\textwidth]{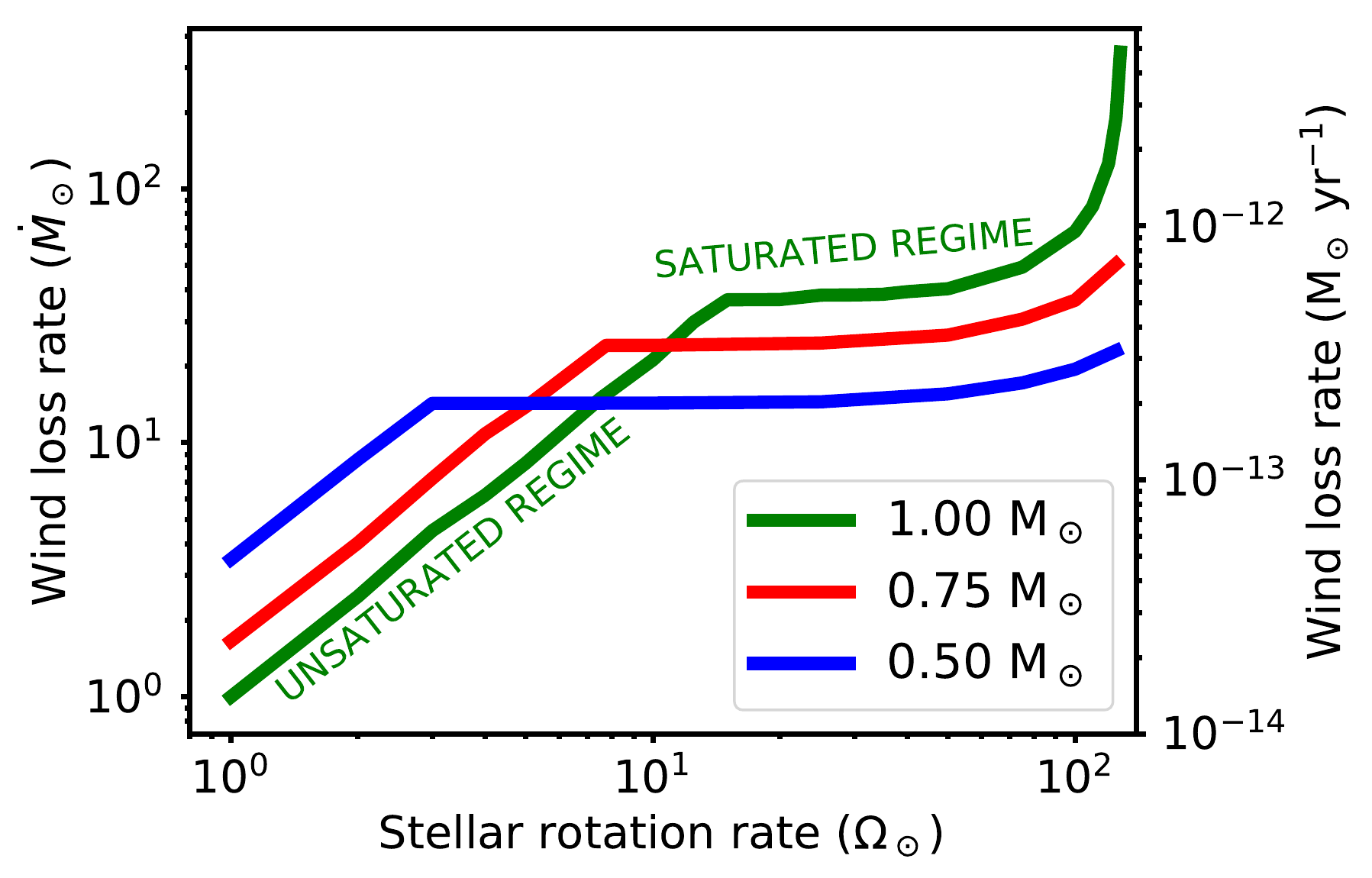}
\vspace{-5mm}
\caption{
\emph{
Figure showing a plausible scenario for the dependence of wind mass loss on rotation for several stellar masses based on the 1D MHD wind model described by Johnstone (\cite{Johnstone17}). 
The mass loss rates for solar mass stars increase rapidly on the right side of the figure as the line approaches the break-up rotation rate.
}
}
\vspace{-4mm}
\label{fig:windmdotrot}
\end{figure}

\subsection{Winds of active stars}

As discussed earlier, little is known about the properties of the winds of active stars largely due to our lack of a theoretical understanding of the underlying mechanisms that heat and accelerate the solar wind and our lack of observational constraints on wind properties.
While many models have been developed to predict the properties of active star winds, we have no clear way to assess the validity of the assumptions made or to test the predictions of the models.
In fact, it is reasonable to wonder if we know anything about the winds of active stars other than that they are likely to have outflow speeds that are within a factor of a few of the solar wind speed and mass loss rates that are within a few orders of magnitude of the modern Sun's mass loss rate. 
Given the observed rotational evolution of rapidly rotating stars, we have good reason to believe that more active stars have higher mass loss rates and this saturates at approximately the same rotation rate that X-ray emission saturates.
A plausible scenario for wind mass loss as a function of rotation is shown in Fig.~\ref{fig:windmdotrot}, where the increase in loss rate for very rapid rotation is discussed below.

The coronae of more active stars are hotter than those of less active stars.
Since both the Sun's X-ray emitting corona and the solar wind are heated by energy transferred from the photosphere by the magnetic field, we might expect that more active stars have hotter winds.
This is plausible and would mean that the wind speeds and mass loss rates are higher for active stars and decay on the main-sequence due to rotational spin-down (Holzwarth \& Jardine \cite{HolzwarthJardine07}; Johnstone et al \cite{Johnstone15b}).
However, while the link between the X-ray emitting coronae and winds of active stars is plausible, several issues exist.
It could be that the mechanisms responsible for heating winds and coronae are different (Cranmer \& van Ballegooijen \cite{CranmervanB10}); for example, it could be that the corona on small scales is be heated by magnetic reconnection while the wind is heated by the dissipation of Alfv\'en waves.
If true, these different mechanisms could respond to changes in a star's magnetic field differently. 
The approximately constant mass loss rate of the Sun over the cycle despite large changes in the X-ray luminosity makes the link between coronal activity and winds unclear (Cohen \cite{Cohen11}) though we should be cautious with interpreting this lack of correlation since the spin-down related decay in stellar magnetic fields and X-ray emission is not analogous to the changes in the Sun's magnetic field and X-ray emission over the solar cycle.
Even if the winds of more active stars are hotter, Suzuki et al. (\cite{Suzuki13}) suggested that the higher densities in the wind cause enhanced radiative cooling that can reduce the final speeds and saturate the mass loss.

For very rapidly rotating stars, other processes become important.
Although the third and forth terms on the right hand side of Eqn.~\ref{eqn:wdmodel} for the centrifugal and Lorentz forces respectively are negligible for the Sun, they are significant for the winds of more strongly magnetized and more rapidly rotating stars.
This was studied by Belcher \& MacGregor (\cite{BelcherMacGregor76}) who identified two distinct regimes.
Stars are in the slow magnetic rotator regime when centrifugal and Lorentz forces have a negligible contribution to the acceleration of the wind. 
Stars are in the fast magnetic rotator regime when these forces dominate in determining the flow speeds of the wind far from the star.
In this regime, the mass loss rate is still determined primarily by other processes.
It is also possible for very rapidly rotating stars to be in a third regime, called the centrifugal magnetic rotator regime, where centrifugal and Lorentz forces influence significantly the wind mass loss rates.
A star enters this regime when the equatorial corotation radius is below the sonic point. 
The properties of winds for stars that are very rapidly rotating were studied by Johnstone (\cite{Johnstone17}) who showed that as stars approach the break-up rotation rate, wind mass loss rates rapidly increase as shown in Fig.~\ref{fig:windmdotrot}, though they also found that the wind properties in the fast and centrifugal rotator regimes depend on assumptions made for the wind temperatures of active stars.

\subsection{Coronal mass ejections} \label{sect:cmes}

There is another factor that complicates the picture and adds another level of uncertainty to our understanding of the winds of active stars.
Coronal mass ejections (CMEs) are eruptive events involving a bulk of plasma with its magnetic field that is accelerated away from the Sun and propagates through interplanetary space. 
Solar CMEs have a very large range of properties and typically have speeds between 100 and 1000~km~s$^{-1}$ and masses between $10^{13}$ and \mbox{$5 \times 10^{16}$}~g (Aarnio et al. \cite{Aarnio11}).
Their angular widths can be as small as 10$^\circ$ or larger than 180$^\circ$ (Yashiro et al. \cite{Yashiro04}).
The rate at which the Sun produces CMEs varies with the solar cycle, with a higher CME occurrence rate being seen at solar maximum (Webb \& Howard \cite{WebbHoward94}).
There is a clear link between CMEs and solar flares (Harrison \cite{Harrison95}; Aarnio et al. \cite{Aarnio11}) and it is likely that in many cases, flares and CMEs result from the same magnetic events, but the exact nature and universality of this link are unclear. 
While flares and CMEs often occur together, many CMEs are observed without a corresponding flare and many flares are observed without a corresponding CME, though this is surely influenced by difficulties in detecting and linking these two phenomena.
It appears that the link with flares is stronger for larger solar flares, and Wang \& Zhang (\cite{WangZhang07}) estimated that 90\% of X-class flares (the largest class of flares) are associated with CMEs. 
More energetic flares are associated with more massive and rapidly outflowing CMEs.
For a review of solar CMEs, see Chen~\cite{Chen11}.

Since more active stars have higher flare occurrence rates, we should expect that they have also higher CME rates. 
This is difficult to study since, like winds, CMEs are very difficult to detect observationally on other stars (Leitzinger et al. \cite{Leitzinger14}) though some possible detections exist (Moschou et al. \cite{Moschou19}; Argiroffi et al. \cite{Argiroffi19}; Leitzinger et al. \cite{Leitzinger20}).
One way to study CMEs from active stars is to combine statistical relations for the solar flare--CME relation with observationally determined statistics for flare rates and energies on active stars.
Aarnio et al. (\cite{Aarnio13}) studied CME winds for pre-main-sequence stars and estimates mass loss rates of between $10^{-12}$ and $10^{-9}$~M$_\odot$~yr$^{-1}$. 
For saturated Sun-like main-sequence stars, Drake et al. (\cite{Drake13}) estimated CME mass loss rates of $10^{-10}$~M$_\odot$~yr$^{-1}$.
For a review of these types of extrapolations, see Osten \& Wolk (\cite{OstenWolk15}).
These high CME rates would lead to significant angular momentum removal and could be one of the factors influencing the rotational evolution of rapidly rotating stars (Aarnio et al. \cite{Aarnio13}; Jardine et al. \cite{Jardine20}).

These results suggest that the winds of very active stars could be entirely dominated by CMEs and planets orbiting such stars could experience highly turbulent space weather conditions.
However, several issues exist with these estimates.
The high mass loss estimated by Drake et al. (\cite{Drake13}) leads to a continuous CME energy loss that is 10\% of the star's bolometric luminosity which is unreasonably large.
They suggested that the flare--CME relation must be invalid for the largest flares to avoid such extreme energy loss.
A possible solution to this problem is the suppression of stellar CMEs by the global magnetic fields of host stars (Alvarado-G\'omez et al. \cite{AlvaradoG18}).
While the global component of the Sun's magnetic field is typically $\sim$1~G, young active Sun-like stars have values that are orders of magnitude larger (Folsom et al. \cite{Folsom16}) and these very strong fields can trap CMEs within the closed corona.

%===========================================================================================================

\section{Earth's Upper Atmosphere and Wind Interactions}

To understand the diverse ways that stellar winds can influence the atmospheres of planets, it is useful to consider first the modern Earth's since this is the most well understood case. 
The modern Earth's atmosphere is composed primarily of N$_2$ and O$_2$ (and 1\% Ar which is of little important), with small amounts of other species such as CO$_2$ and H$_2$O, and the Sun's magnetic activity is generally very low. 
In this section, I describe the structure and physics of the Earth's upper atmosphere and magnetosphere, and in the next section, I describe how this is likely to be different for planets with different types of atmospheres orbiting different types of stars. 

\subsection{The vertical structure of the Earth's atmosphere} \label{sect:verticalat}

\begin{figure}
\centering
\fbox{\includegraphics[trim = 0mm 0mm 0mm 0mm, clip=true,width=0.7\textwidth]{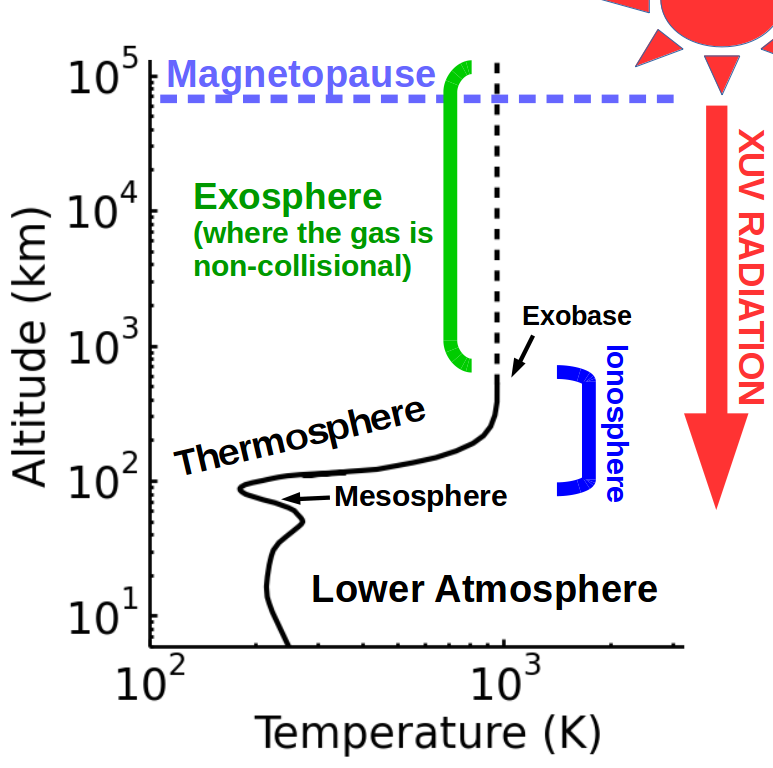}}
\vspace{-0mm}
\caption{
\emph{
Cartoon showing the vertical structure of Earth's atmosphere.
}
}
\vspace{-5mm}
\label{fig:earthlayers}
\end{figure}

It is common to break down the vertical structure of the atmosphere is into layers based on the temperature gradient as shown in Fig.~\ref{fig:earthlayers}. 
The lowest layer is the troposphere which extends from the surface to the tropopause at an altitude of 10~km and has a decreasing temperature with increasing altitude (i.e. a negative temperature gradient). 
The troposphere is heated by its contact with the surface and this heat is distributed upwards by convection. 
Above the troposphere is the stratosphere, which has a positive temperature gradient due to the absorption of solar ultraviolet radiation by ozone and extends from 10~km to the stratopause at approximately 50~km. 
These two regions are labeled in Fig.~\ref{fig:earthlayers} as the `lower atmosphere'. 
Above this is the mesosphere, which again has a negative temperature gradient due to cooling by the emission of infrared radiation to space by CO$_2$ molecules and extends from 50~km to the mesopause at approximately 100~km. 
Above this is the thermosphere, which has a strong positive temperature gradient due to the absorption of solar X-ray and extreme and far ultraviolet radiation, particularly by O$_2$ and O.
The temperature of the Earth's thermosphere typically varies between approximately 500 and 1500~K and depends primarily on the Sun's activity (Roble et al. \cite{Roble87}).
The thermosphere extends to the exobase, which is typically at an altitude of a few hundred km and varies also with the Sun's activity. 
The exobase is where the gas density drops low enough that the gas is approximately non-collisional.

\begin{figure}
\centering
\includegraphics[trim = 0mm 0mm 0mm 0mm, clip=true,width=0.32\textwidth]{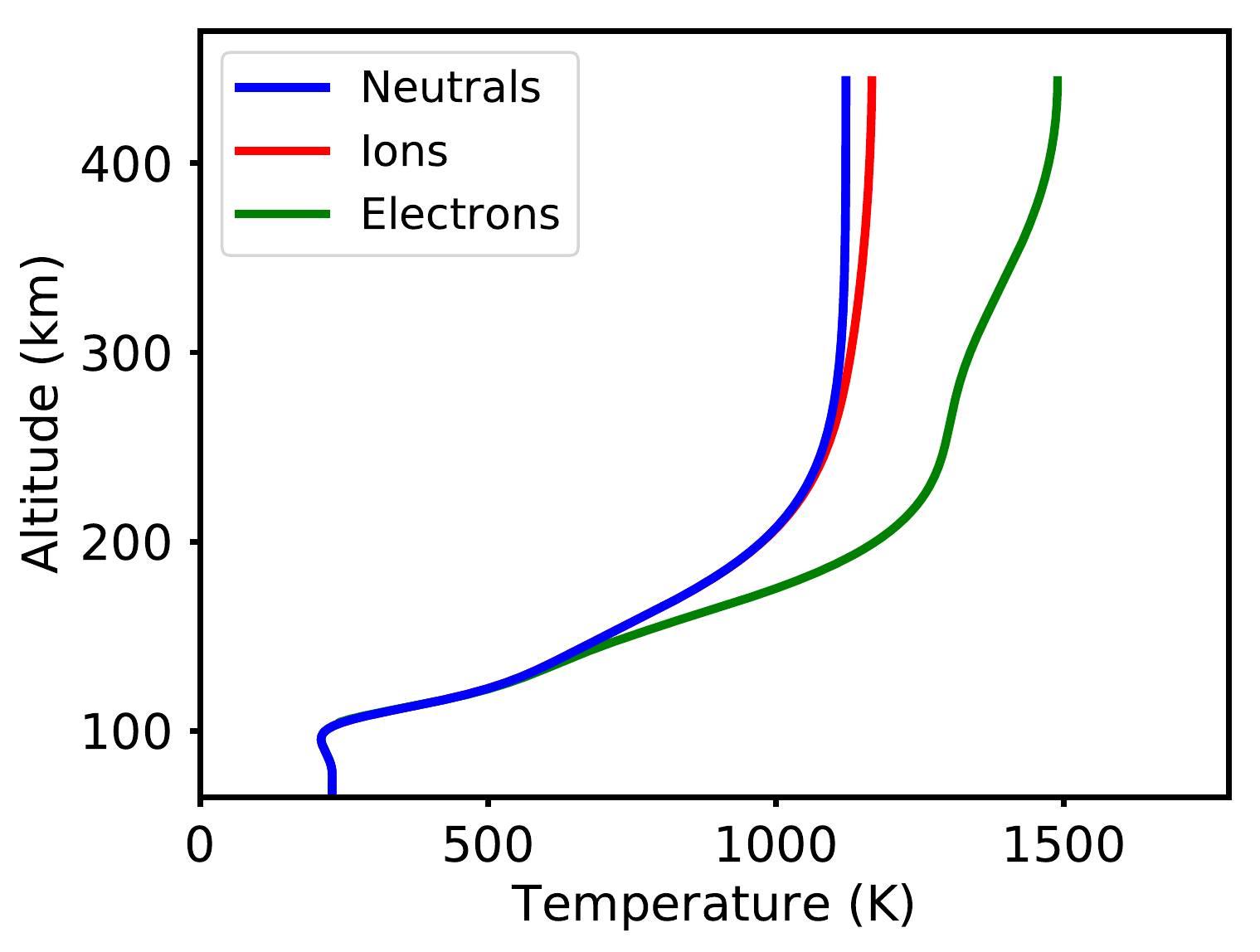}
\includegraphics[trim = 0mm 0mm 0mm 0mm, clip=true,width=0.32\textwidth]{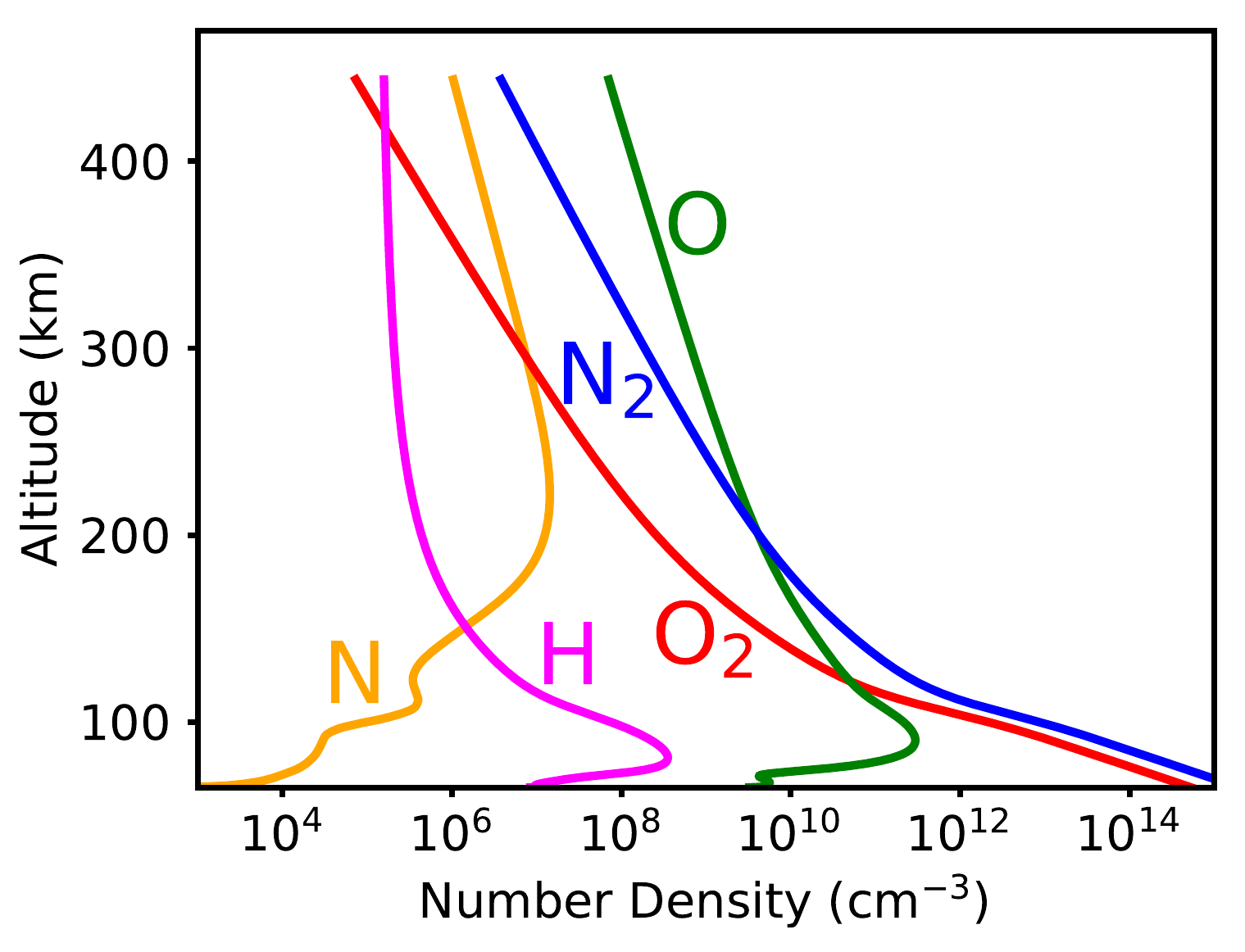}
\includegraphics[trim = 0mm 0mm 0mm 0mm, clip=true,width=0.32\textwidth]{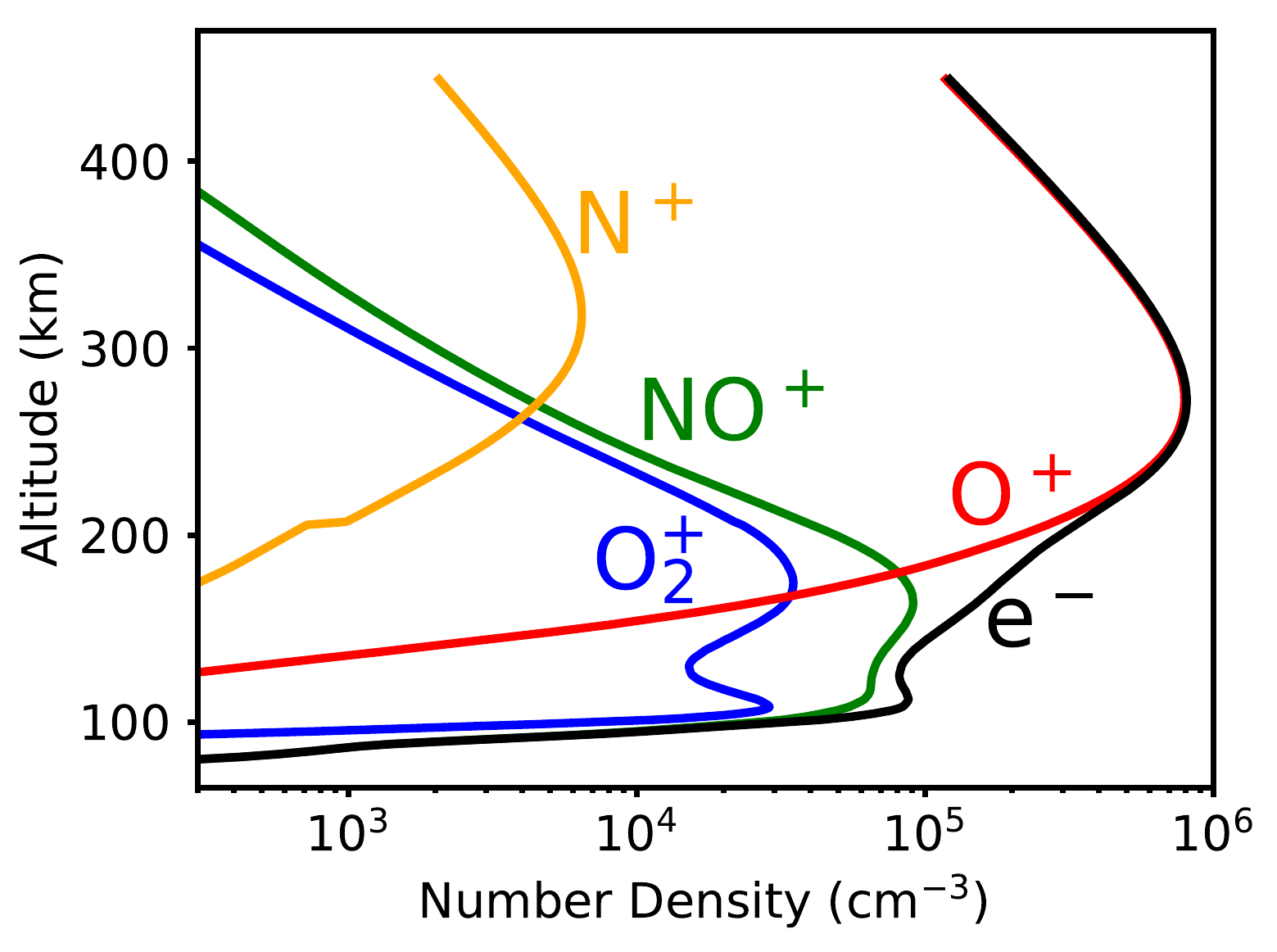}
\vspace{-0mm}
\caption{
\emph{
The temperature and chemical structure of the Earth's upper atmosphere as calculated by Johnstone et al. (\cite{Johnstone18}).
}
}
\vspace{-0mm}
\label{fig:earththermosphere}
\end{figure}

It is useful also to consider the chemical structure of the atmosphere which also has several layers, the most important of which are the homosphere and the heterosphere.
The homosphere extends from the surface until approximately 120~km and in this region the gas is well mixed by turbulent processes, causing the mixing ratios of stable long-lived species (e.g. N$_2$, O$_2$, CO$_2$, Ar, etc.) to be uniform with altitude regardless of their masses.
An important exception to this is water vapor which has a very low abundance at high altitudes due to the presence of the cold trap in the lower stratosphere which prevents water molecules from reaching the upper atmosphere. 
The upper boundary of the homosphere is called the homopause or turbopause. 
Above the homopause is the heterosphere and in this region molecular diffusion separates the chemical species by mass such that heavier species become increasingly less abundant at higher altitudes.
The Sun's XUV radiation dissociates molecules and causes the gas throughout most of the thermosphere to be dominated by atomic O and N as shown in Fig.~\ref{fig:earththermosphere}.  

If we consider a hydrostatic atmosphere only, the density of the $i$th chemical species, $n_i$, decreases with altitude, $z$, as \mbox{$n_i \propto \exp \left( - z / H_i \right)$}, where $H_i$ is the pressure scale height given by \mbox{$k T / m g$}, where $k$ is the Boltzmann constant, $T$ is temperature, $m$ is the molecular mass of the gas, and $g$ is the gravitational acceleration. 
Note that this only applies to long lived species that are not rapidly created and destroyed by chemical processes.
The difference between the homosphere and the heterosphere is the molecular mass term, $m$.
In the homosphere, this is the average molecular mass of the entire gas meaning that each chemical species has the same scale height regardless of mass.
In the heterosphere, this is the mass of individual species, meaning that heavier species have smaller scale heights.

The upper mesosphere and thermosphere correspond to the ionosphere where the absorption of solar X-ray and EUV photons causes ionisation of the gas.
The density of ions depends sensitively on the Sun's activity and varies over the solar cycle (e.g. Solomon et al. \cite{Solomon13}).
For the modern Earth, the ionisation fraction of the gas reaches values of $\sim 10^{-3}$ at typical solar maximum conditions and this small mixing of ions is very important for interactions between the upper atmosphere and the magnetosphere. 
By definition the ionosphere extends to the exobase and since the gas densities are low in the upper thermosphere, the neutral, ion, and electron components of the gas have different temperatures.
The thermal and chemical structure of the ionosphere is shown in Fig.~\ref{fig:earththermosphere}.

\subsection{Exospheres, magnetospheres, and winds}

\begin{figure}
\centering
\includegraphics[trim = 0mm 0mm 0mm 0mm, clip=true,width=0.95\textwidth]{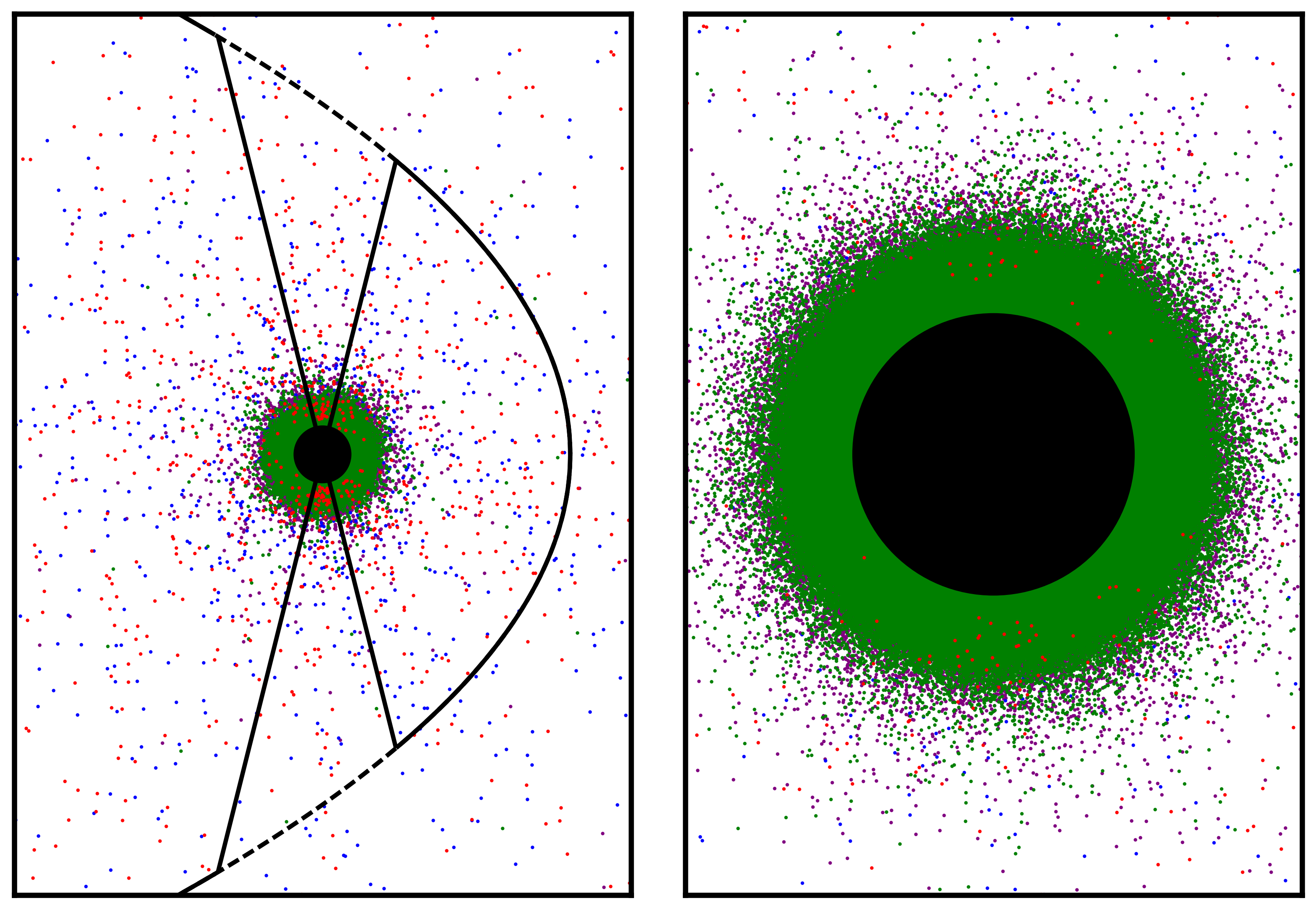}
\vspace{-0mm}
\caption{
\emph{
The basic structure of the exosphere of the Earth assuming the Sun's activity from 3~Gyr ago as calculated using a kinetic particle model by Kislyakova et al. (submitted), with each point corresponding to approximately $10^{29}$~particles.
The right panel shows the same as the left panel but zoomed in around the planet which is shown in black.
The purple and green points show show atomic nitrogen and oxygen neutrals respectively, and the blue and red points show atomic nitrogen and oxygen ions.
The black parabola in the left panel shows the approximate boundaries of the magnetosphere and the wedge shaped lines coming from the poles show the polar opening angle expected 3~Gyr ago due to stronger solar wind conditions.
Image courtesy of K. G. Kisylakova.
}
}
\vspace{-0mm}
\label{fig:exoearth}
\end{figure}

Above the exobase is the exosphere, which is approximately isothermal and extends out into interplanetary space (there is no clear and exact definition for the upper boundary of the exosphere).
A useful definition for the exobase is based on the Knudsen number, $Kn$, which is a dimensionless quantity given by the ratio of the mean free path of particles to a relevant length that characterises the scale of the system, which for planetary atmospheres is typically the pressure scale height, which we can define as \mbox{$H = - p / (dp/dz)$}, where $p$ is the thermal pressure and $z$ is the altitude or radius, and for a fully hydrostatic atmosphere this is given by \mbox{$k T / m g$}.
Low in the Earth's atmosphere, the high density of the gas means \mbox{$Kn \ll 1$} and as we go to higher altitudes this value increases until the exobase where it becomes unity.
Below the exobase, the particle speed distribution can be approximated by a normal thermal Maxwellian distribution.
In the exosphere, the much more tenuous gas means that \mbox{$Kn \gg 1$} and particles are mostly able to move freely with few collisions and they do not follow the Maxwellian distribution.
A particle simulation of the exosphere of the Earth, assuming solar conditions from 3~Gyr ago, is shown in Fig.~\ref{fig:exoearth}.

Particles that cross the exobase can have a range of fates depending on their velocities and their interactions with other exospheric particles (Beth et al. \cite{Beth14}) and with the solar radiation field and wind (Zhang et al. \cite{Zhang93}; Beth et al. \cite{Beth16}). 
In the absence of any interactions, particles either fall back to the atmosphere on ballistic trajectories or escape the system entirely depending on if they are traveling slower or faster than the escape velocity.
If they collide with other exospheric particles, they can become satellite particles that orbit the planet.  
Particles can also become ionised by absorbing solar XUV photons, by colliding with electrons in the solar wind, or by charge exchanging with ions in the solar wind. 
If ionised, their trajectories are also influenced significantly by the Earth's magnetosphere or by the solar wind depending on where the ionisation takes place, with ions created beyond the boundary of the magnetosphere being swept away by the solar wind magnetic field.
As discussed in Section~\ref{sect:windobsplanet}, charge exchange between exospheric particles and wind protons creates a population of fast moving neutrals called ENAs.
Since these ENAs are not influenced by the planet's magnetosphere, ENAs created between the planet and the direction from which the wind is propagating enter the atmosphere and collide with atmospheric particles, which can heat the upper atmosphere (Chassefi{\`e}re \cite{Chassefiere97}; Lichtenegger et al. \cite{Lichtenegger16}).

When considering atmospheric loss processes, an important source of ions in the magnetosphere is upward flows from the ionosphere.
Given their much lighter mass, ionospheric free electrons are poorly bound gravitationally and expand away from the planet causing a reduction in the electron density and the generation of an ambipolar electric field.
This field acts to maintain quasi-neutrality in the gas by inhibiting the outflow of electrons, which causes an outward pressure on the ions resulting in a flow of ions and electrons from the ionosphere into the magnetosphere.
These particles are then distributed throughout the magnetosphere (Ebihara et al. \cite{Ebihara06}) and their ultimate fates are currently a matter of debate.
While some studies suggest that most escape the magnetosphere and are lost to interplanetary space (Nilsson et al. \cite{Nilsson12}), others suggest that most are unable to escape and are recaptured by the planet (Seki et al. \cite{Seki01}).
The outward flow of ions and the fraction of these ions that escape to space depend on the solar wind conditions and emission of high-energy radiation (Haaland et al. \cite{Haaland12}).
For a review of ion flows in the magnetosphere, see Moore \& Horwitz (\cite{MooreHorwitz07}).

\begin{figure}
\centering
\includegraphics[width=0.85\textwidth]{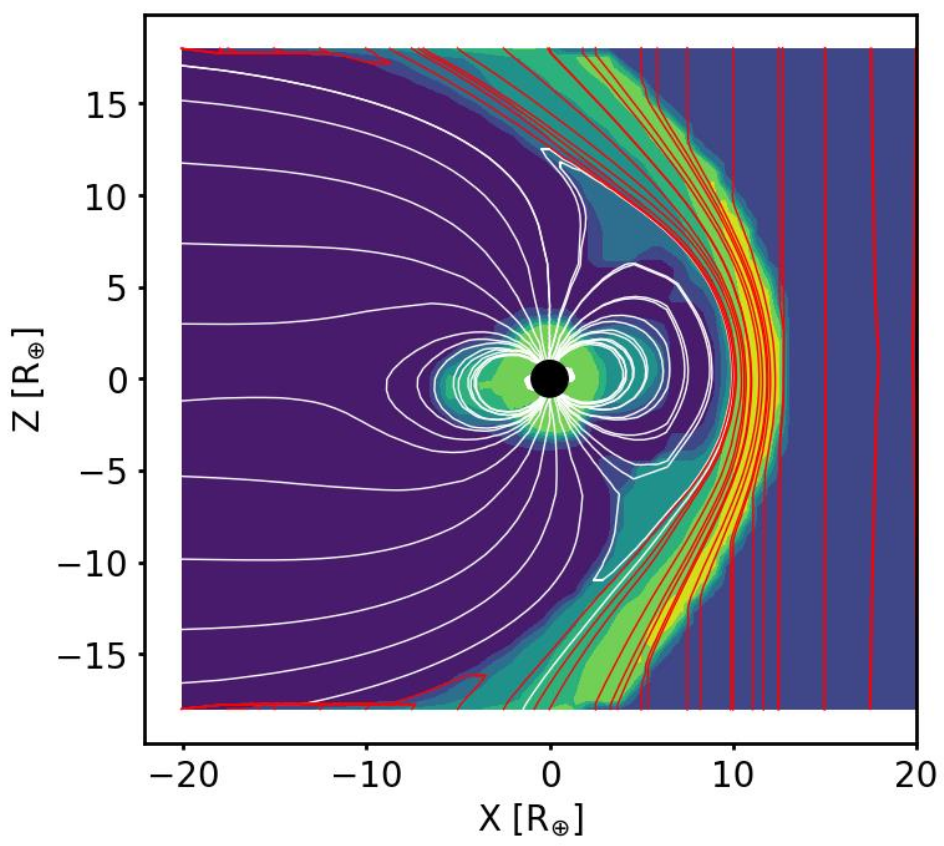}
\vspace{-5mm}
\caption{
\emph{
Simulation of Earth exposed to dense and fast early solar wind calculated using the Space Weather Modelling Framework (T\'oth et al. \cite{Toth05}). 
The background color shows the plasma density, and red and white lines illustrate the planetary and solar wind magnetic field lines.
Image courtesy of K. G. Kislyakova.
}
}
\vspace{-0mm}
\label{fig:magnetosphere}
\end{figure}

The structure of Earth's magnetosphere is shown in Fig.~\ref{fig:magnetosphere}.
For the Earth, the boundaries of the magnetosphere are very large in comparison to the planet and the atmosphere.
The outer boundary of the magnetosphere is called the magnetopause and in the direction of the incoming solar wind is approximately at the point where the magnetic pressure from the Earth's magnetic field balances the ram pressure in the wind, which is typically at a radius of $\sim$10~R$_\oplus$.
Since the exobase is typically at a radius of 1.07~R$_\oplus$, the very large magnetosphere compared to the thickness of the atmosphere means that most exospheric particles do not interact directly with the solar wind.
For this reason, the magnetosphere is often described as protecting the atmosphere from escape to space by wind erosion. 
However, the situation is more complicated and there is disagreement about the effects of the magnetosphere on escape, as discussed below. 
Note that planets that do not possess significant intrinsic magnetic fields, such as Venus and Mars, form induced magnetospheres due to interactions between the solar wind and their ionospheres and exospheres (Bertucci et al. \cite{Bertucci11}).
The interaction between the solar wind and the Earth's magnetosphere also causes additional effects lower in the atmosphere, especially during geomagnetic storms when the rapid changes in the space weather conditions in Earth's vicinity causes reconnection in the magnetosphere and jostling of the magnetic field lines that connect to the Earth's ionosphere
(Chappell \cite{Chappell16}).
The former causes the acceleration of particles that can precipitated downwards into the thermosphere and below, driving thermal and chemical processes.
The latter causes electric currents in the ionosphere, and as these charged particles collide with neutrals, energy is dissipated causing the gas to be heated (this process is called Joule heating), especially in the region abouve 100~km (Roble et al. \cite{Roble87}).

\section{Planetary Upper Atmospheres and Escape}

Loss processes take place in the planet's thermosphere and exosphere and are mostly driven by the central star's output of high-energy radiation and wind.
These processes are often broken down into thermal and non-thermal processes (Lammer 2013) where the thermal processes are Jeans escape and hydrodynamic escape and non-thermal processes include everything else. 
Jeans escape involves the high energy tail of the thermal Maxwellian distribution at the exobase with upward speeds greater than the planet's escape velocity, allowing the particles to leave the planet entirely.
This is especially important for low mass particles such as hydrogen, but can also be important for higher higher mass particles if the atmosphere is very hot.
Several processes in the planet's exosphere ionize neutral particles leading them to be picked up by the stellar wind (Kislyakova et al. \cite{Kislyakova14a}; Bourrier et al. \cite{Bourrier16}).
Other significant processes include the outflow of ions from the magnetic poles as discussed above (Glocer et al. \cite{Glocer09}) and photochemical escape, which involves the escape high speed particles created in chemical reactions (Amerstorfer et al. \cite{Amerstorfer17}) and is powered by energy absorbed from the star's XUV spectrum.
In all these cases, the loss rates are influenced heavily by the chemical and thermal structure of the upper atmosphere, with hotter atmospheres having higher loss rates, in some cases because they are more expanded and therefore more exposed to direct interactions with the star's wind. 
When gas temperatures are very high, it is possible for the upper atmosphere to flow away from the planet hydrodynamically (e.g. Tian et al. \cite{Tian05}) at very high rates. 
The structure of the upper atmosphere is determined by many thermal and chemical processes and is influenced by several factors, including the mass of the planet, the chemical composition of the atmosphere, and the star's X-ray and ultraviolet spectrum.

It is important to understand what effect an intrinsic planetary magnetic field has on these escape processes and on the total escape.
It is often assumed that by reducing the direct interactions between a planet's exosphere and the star's wind, magnetic fields protect atmospheres from escape.
However, there are a large number of loss processes that respond to planetary magnetic fields in complex ways and the net effect of a magnetosphere is not clear.
While it is true that direct ionisation and pickup of exospheric neutral particles by winds is prevented by magnetospheric shielding, the magnetosphere collects energy from the wind and dissipates this energy lower in the atmosphere, driving processes such as high energy particle acceleration and Joule heating.
Gunell et al. (\cite{Gunell18}) studied the importance of the presence and strength of a planet's intrinsic magnetic field on several loss processes and found that in some cases, the magnetic field increases the escape of ions.  
It is possible that given the complexity of the problem, a planetary magnetic field influences escape in different ways for different parts of the parameter space. 
For example, this was found by Egan et al. (\cite{Egan19}) who studied ion escape from weakly-ionized Mars-sized planets.

Gunell et al. (\cite{Gunell18}) pointed out that the magnetised Earth and the unmagnetised Venus and Mars all lose atmosphere at approximately the same rate ($\sim$0.5 to 2 kg~s$^{-1}$).
While this is suggestive, differences in the masses, orbital distances, and atmospheric compositions of these planets possibly explain the similar loss rates.
For example, while Venus is closer to the Sun and lower mass than Earth, its atmosphere is composed mostly of carbon dioxide which, as I explain below, cools the upper atmosphere and protects it from escape, whereas Earth's N$_2$ and O$_2$ dominated atmosphere heats up and expands much more readily.
Mars is further from the Sun than Earth and also has a carbon dioxide atmosphere, which reduces the escape, but it also has a very low mass.

\subsection{Hydrodynamic escape}

An important loss mechanism that has been discussed extensively in the exoplanet literature is hydrodynamic escape.
The term is usually used to refer to the outflow of atmospheres that are heated to such an extent that they flow away from the planets in the form of transonic Parker winds.
This process does not take place in the solar system, mostly due to the Sun's low activity.
Hydrodynamic escape can be driven by the XUV spectrum of active stars (e.g. Tian et al. \cite{Tian05}), the energy held within the planet and atmosphere immediately after the circumstellar gas disk dissipates (St\"okl et al. \cite{Stoekl15}), or the photospheric emission of the star for very short period planets (Owen \& Wu \cite{OwenWu16}).
Since atmospheres composed of lighter gases are lost more rapidly (Erkaev et al. \cite{Erkaev13}), these processes are most important for hydrogen dominated primordial atmospheres.
For terrestrial planets with such atmospheres orbiting active stars, hydrodynamic escape is much stronger than stellar wind stripping (Kislyakova et al. \cite{Kislyakova13}).
It is even possible that the pressure of the star's wind suppresses the outflow and reduces the escape (Shaikhislamov et al. \cite{Shaikhislamov16}; Vidotto \& Cleary \cite{VidottoCleary20}).
In hydrodynamically outflowing atmospheres, adiabatic cooling driven by the expansion of the gas has been shown to be an important thermal process (Tian et al. \cite{Tian05}; Murray-Clay et al. \cite{MurrayClay09}).

A simple estimate for XUV-driven hydrodynamic escape rate can be achieved by considering the amount of energy in the star's XUV field that is available to remove gas from the planet. 
If we assume a planet absorbs all XUV radiation that passes within a radius $R_\mathrm{XUV}$ of the planet's center, the rate at which the planet absorbs stellar radiation is given by \mbox{$\pi R_\mathrm{XUV}^2 F_\mathrm{XUV}$}, where $F_\mathrm{XUV}$ is the XUV flux at the planet's orbit.
It is useful to define a parameter $\epsilon$ as the fraction of absorbed energy used to lift mass out of the gravitational potential well of the planet.
This term is often called the `heating efficiency' and this has caused it to be confused with the fraction of absorbed XUV energy that is used to heat the gas.
Much of the absorbed energy that is released as heat is irradiated back to space or ends up as kinetic energy of the escaping escaping gas which is typically accelerated to speeds beyond that needed to leave the system.
Since the energy required to remove a gravitationally bound mass $m$ from near the surface of a planet is given by \mbox{$G M_\mathrm{pl} m / R_\mathrm{pl}$}, the rate at which a hydrodynamic flow removes energy is \mbox{$G M_\mathrm{pl} / R_\mathrm{pl} \dot{M}_\mathrm{at}$}, where $\dot{M}_\mathrm{at}$ is the atmosphere's mass loss rate.
This leads to an estimate of the mass loss rate given by
\begin{equation}
\dot{M}_\mathrm{at} = \frac{\epsilon \pi R_\mathrm{XUV}^2 R_\mathrm{pl} F_\mathrm{XUV}}{G M_\mathrm{pl}}
\end{equation}
which is typically called the energy-limited escape equation (Watson et al. \cite{Watson81}).

This simple reasoning does not take into account several factors that can influence hydrodynamic losses. 
For instance, tidal effects are important for close-in planets (Erkaev et al \cite{Erkaev07}).
Since some of the energy is absorbed in the supersonic part of the wind and therefore cannot contribute to the loss rate and since the sonic point is closer to the planet for more rapidly outflowing atmospheres, the dependence of $\dot{M}_\mathrm{at}$ on $F_\mathrm{XUV}$ should be weaker than linear (see Fig.~1 of Johnstone et al. \cite{Johnstone15letter}). 
Similarly, the loss of absorbed energy by radiative processes causes a weaker than linear $\dot{M}_\mathrm{at}$--$F_\mathrm{XUV}$ dependence for very highly irradiated planets (Murray-Clay et al. \cite{MurrayClay09}; Owen \& Jackson \cite{OwenJackson12}).
Also important for $F_\mathrm{XUV}$ is which parts of the XUV spectrum contribute to escape and this depends on the chemical composition of the atmosphere since the wavelength dependence for absorption are different for different chemical species.
Since H and H$_2$ do not efficiently absorb radiation longward of 112~nm, only the spectrum shortward of this value is important.
However, for Earth-like atmospheres, absorption by species such as O$_2$ and O$_3$ means that the relevant part of the spectrum extends beyond 200~nm, meaning more energy is available to drive hydrodynamic escape (Johnstone et al. \cite{Johnstone19a}).
For very strongly escaping hydrogen dominated atmospheres, it is possible that only the X-ray part of the spectrum is relevant if the EUV radiation is entirely absorbed above the sonic point (Owen \& Jackson \cite{OwenJackson12}).
For a description of hydrodynamic escape that overcomes much of the limitations of the energy limited equation, see Kubyshkina et al. (\cite{Kubyshkina18a}, \cite{Kubyshkina18b}).

\begin{figure}
\centering
\includegraphics[trim = 0mm 0mm 0mm 0mm, clip=true,width=0.85\textwidth]{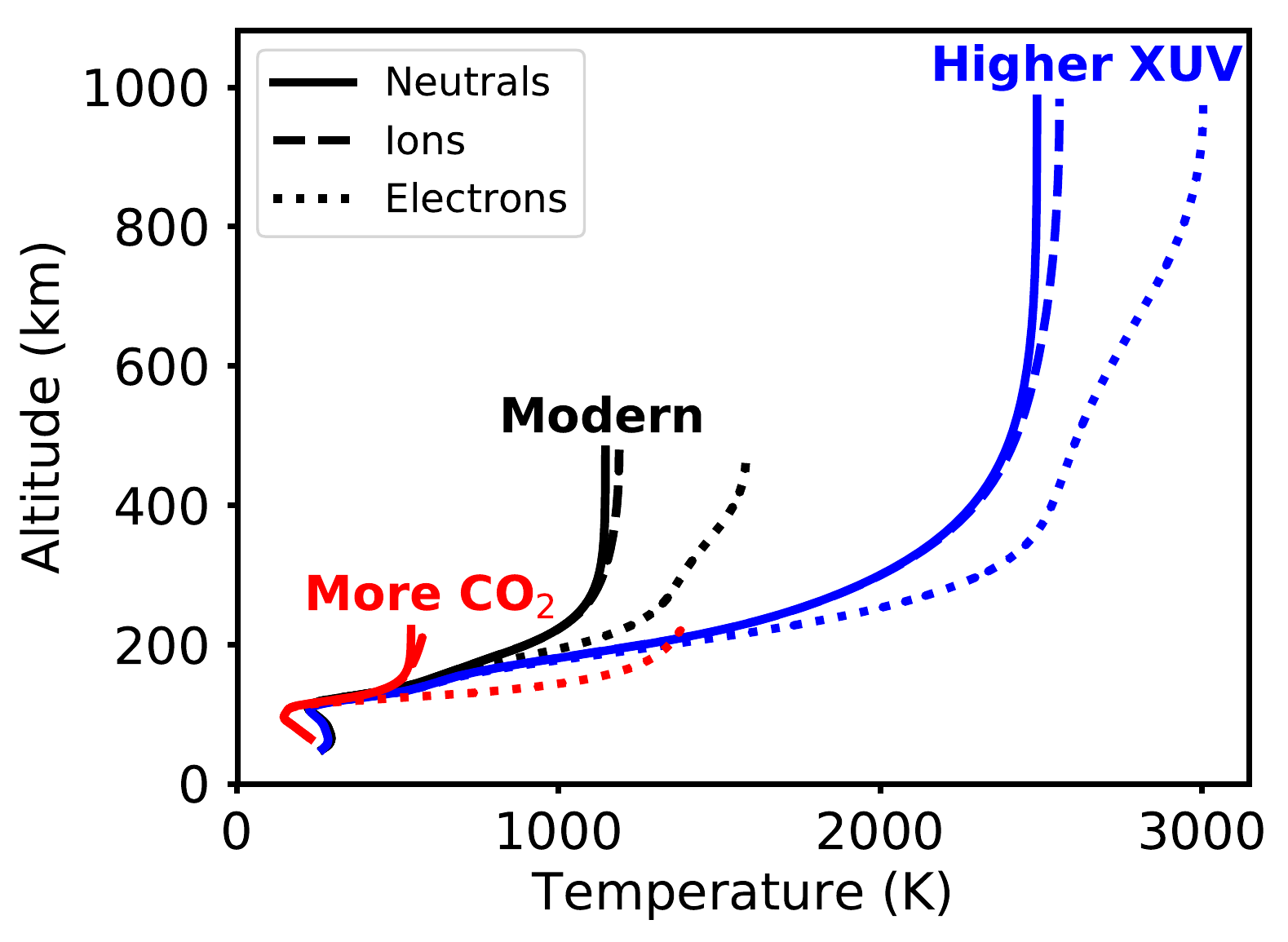}
\vspace{-5mm}
\caption{
\emph{
Thermal structure of Earth's mesosphere and thermosphere assuming modern conditions (black lines), the modern Earth's atmosphere exposed to the higher XUV flux of the Sun approximately 2.5~Gyr ago (blue lines), and an Earth-like atmosphere with 100 times the present CO$_2$ composition exposed to the modern solar spectrum (red lines).
The atmosphere models were calculated by Johnstone et al. (\cite{Johnstone18}).
}
}
\vspace{-5mm}
\label{fig:thermoearth}
\end{figure}

The thermal escape of an atmosphere and whether it is undergoing hydrodynamic escape depends primarily on the planet's mass, the mass of the atmospheric particles, and the temperature of the upper atmosphere and is often characterised by the Jeans escape parameter, $\lambda_\mathrm{J}$.
This is defined as the ratio of the potential energy to the thermal energy per particle at the exobase and is given by
\begin{equation}
\lambda_\mathrm{J} = \frac{G M_\mathrm{pl} m }{ k_\mathrm{B} T_\mathrm{exo} R_\mathrm{exo} } ,
\end{equation}
where $m$ is the molecular mass of the gas, $T_\mathrm{exo}$ is the gas temperature, and $R_\mathrm{exo}$ is the exobase radius.
This is also the ratio of the square of the escape velocity (\mbox{$v_\mathrm{esc}^2 = 2 G M_\mathrm{pl} / R_\mathrm{exo}$}) to the square of the most probable thermal velocity (\mbox{$v_\mathrm{th}^2 = 2 k_\mathrm{B} T_\mathrm{exo} / m$}) for particles at the exobase.  
This parameter specifies how gravitationally bound an atmosphere is to the planet and for $\lambda_\mathrm{J} > 30$ the atmosphere can safely assumed to be hydrostatic, and for $\lambda_\mathrm{J} < 15$, rapid thermal escape is expected to take place. 
For considering the escape of individual chemical species, $m$ should be the mass of that species, and for considering if an atmosphere is undergoing hydrodynamic escape, $m$ should be the average molecular mass of the gas.
The name of this parameter comes from the fact that it is used in the expression for the Jeans escape rate, which for given species is given by 
\begin{equation}
\dot{M}_\mathrm{J} = 4 \pi R_\mathrm{exo}^2 n m v_\mathrm{th} \frac{ \left( 1 + \lambda_\mathrm{J} \right) \mathrm{e}^{-\lambda_\mathrm{J}} }{2 \pi^{1/2}} ,
\end{equation}
where $n$ is the number density at the exobase of the species of interest.
For discussions on these processes, see Luger et al. (\cite{Luger15}) and Fossati et al. (\cite{Fossati17}).

\subsection{Upper atmospheres and active stars}

\begin{figure}
\centering
\includegraphics[trim = 0mm 0mm 0mm 0mm, clip=true,width=0.85\textwidth]{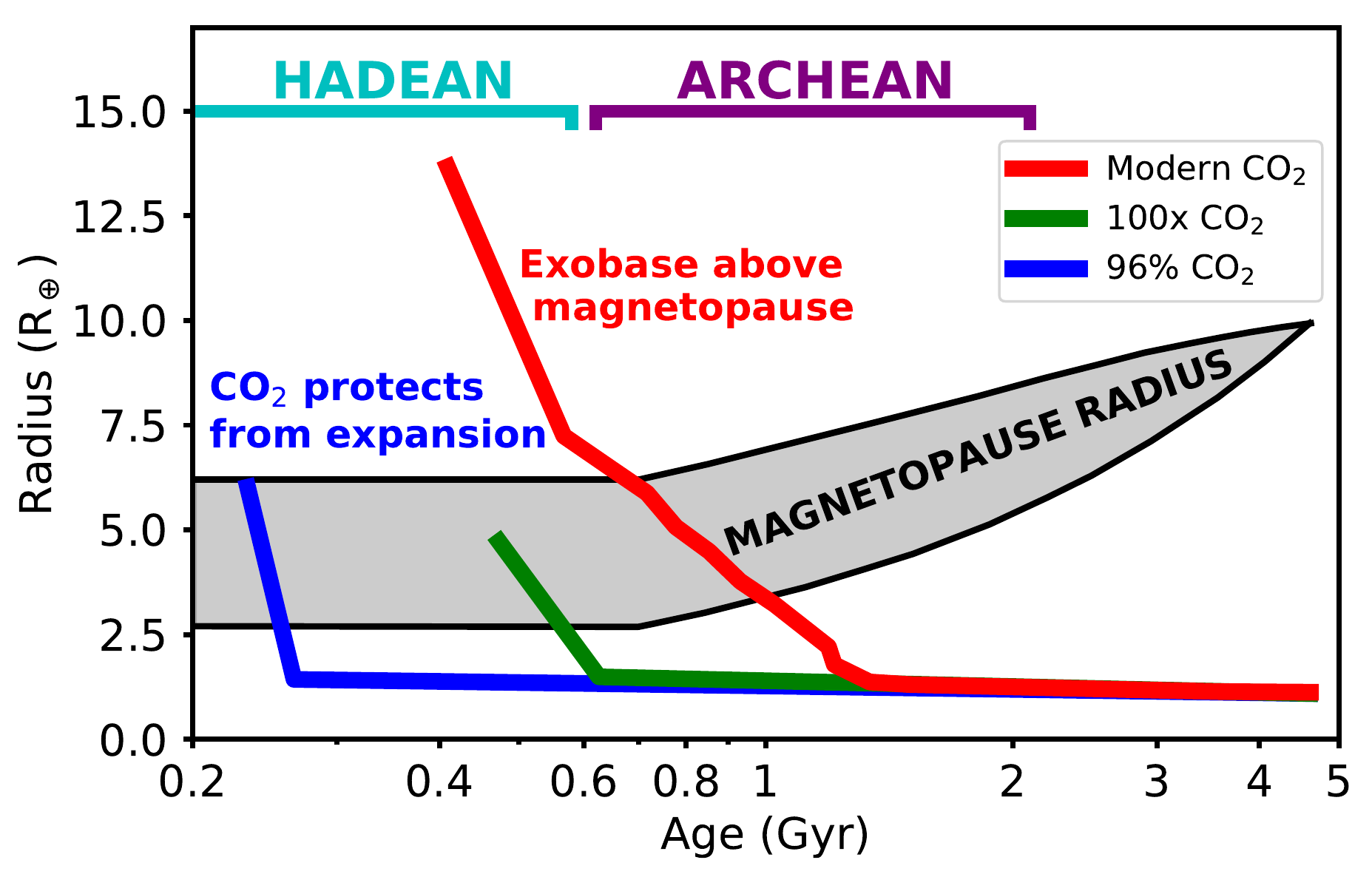}
\vspace{-5mm}
\caption{
\emph{
Estimates from Lichtenegger et al. (\cite{Lichtenegger10}) of the evolution of Earth's exobase altitude between 0.2 and 5~Gyr from the birth of the solar system assuming a modern atmospheric composition (red line), an atmosphere with 100x the modern CO$_2$ (green line), and an atmosphere that is 96\% composed of CO$_2$ (blue line).
The shades area shows their range of estimates for the magnetopause radius.
This figure shows that with a modern atmospheric composition, the Earth's thermosphere would have been directly exposed to the solar wind for the entire Hadean and the early Archean, and this can be avoided if a larger amount of CO$_2$ was present in the atmosphere.
}
}
\vspace{-5mm}
\label{fig:herbert}
\end{figure}

While the modern Earth is a useful case to understand, we are more interested in this reivew in cases involving planets orbiting active stars since this is when the influences of stellar winds and atmospheric loss processes are most important.
The higher XUV fluxes of more active stars causes the gas to be heated to much higher temperatures and to be ionised to a much greater degree. 
Several studies have explored what happens to the thermal and chemical structure the Earth's thermosphere if it were present in its modern form at earlier times in the solar system's history when the Sun was more active (Kulikov et al. \cite{Kulikov07}; Tian et al. \cite{Tian08}; Johnstone et al. \cite{Johnstone18}, \cite{Johnstone19a}). 
These studies have shown that the higher XUV fluxes of the younger more active Sun lead to thermospheres that are significantly hotter and more expanded leading to more rapid escape at the upper boundary.
An example of the effects of a higher XUV flux on the thermal structure of the upper atmosphere is shown in Fig.~\ref{fig:thermoearth}.

While we typically describe the atmospheres of the planets and moons in our solar system as hydrostatic, it is in fact always the case that some material is escaping from the upper atmosphere to space meaning that there is always a net upward flow of material throughout the atmosphere. 
For the Earth, this is typically much less than 1~cm~s$^{-1}$ at the exobase and is negligible.
Tian et al. (\cite{Tian08}) studied the case of an input XUV spectrum with an EUV flux of twenty times the modern value and found that the thermosphere is heated to over $10^4$~K and the exobase expands to an altitude of $10^5$~km.
This puts the exobase higher than the modern magnetopause of the Earth, meaning that the Earth's thermosphere under such conditions would expand beyond the boundaries of the magnetosphere.
Assuming only Jeans escape, they found that this causes upward flow speeds approaching 1~km~s$^{-1}$ which is high enough that adiabatic cooling of the expanding gas causes a negative temperature gradient in the upper thermosphere.
Additional loss mechanisms, such as cold ion outflows and pick-up by stellar winds would cause additional adiabatic cooling in these cases.

Lichtenegger et al. (\cite{Lichtenegger10}) used the results of Tian et al. (\cite{Tian08}) to estimate the evolution of the Earth's thermospheric expansion and compared these with estimates the magnetosphere's size, as summarised in Fig.~\ref{fig:herbert}.
The magnetosphere was likely smaller at earlier evolutionary phases due to stronger compression by the denser and faster early solar wind and a weaker intrinsic magnetic field (Tarduno et al. \cite{Tarduno10}).
The total magnetospheric compression is uncertain and the range of scenarios calculated by Lichtenegger et al. (\cite{Lichtenegger10}) is shown as the shaded area in Fig.~\ref{fig:herbert}.
They showed that with its current atmosphere, the Earth's thermosphere at ages younger than approximately 700~Myr would be expanded beyond the boundaries of the magnetosphere and be exposed to direct interactions with the solar wind. 
For their most active case, they estimates loss rates from ionisation and pick-up of atmospheric particles by the solar wind of 0.1~bar~Myr$^{-1}$, meaning that the Earth's modern atmosphere would be removed in 10~Myr.
Such rapid escape is unreasonable since it would likely imply that the Earth was without an atmosphere throughout the Hadean and early Archean, contrary to evidence for the presence during these times of an atmosphere, liquid surface water, and even life (Mojzsis et al. \cite{Mojzsis96}; Mojzsis et al. \cite{Mojzsis01}; Catling \& Zahnle \cite{CatlingZahnle20}). 

Johnstone et al. (\cite{Johnstone19a}) considered an Earth with an Earth-like atmosphere exposed the much higher XUV flux of a very young rapidly rotating Sun and showed that the atmosphere is heated to such a high temperature that it flows away from the planet hydrodynamically. 
They found mass loss rates of approximately 10~bar~Myr$^{-1}$, which is high enough to remove the entire modern atmosphere of the Earth in 0.1~Myr. 
They also showed that the gas is composed 20\% of ions, which is several orders of magnitude more than the ionisation fraction in modern Earth's ionosphere.

\subsection{The importance of atmospheric cooling}

Consider again the results of Lichtenegger et al. (\cite{Lichtenegger10}) for the case of the Earth's atmosphere exposed to the higher XUV fluxes of the young Sun.
It is interesting to speculate about what could have prevented the unrealistically high mass loss due to pick-up by the young solar wind that they found.
An important process that likely protected Earth's atmosphere during the early period of high solar activity is upper atmosphere cooling by radiative emission to space.
The most likely candidate for this cooling is infrared emission from CO$_2$ which is the most important coolant in Earth's upper atmosphere and is responsible for the formation of the mesosphere.
Geochemical measurements show that the CO$_2$ abundance in Earth's atmosphere during and after the late Archean was orders of magnitude higher than it is in the modern atmosphere (e.g. Sheldon \cite{Sheldon06}; Driese et al. \cite{Driese11}; Kanzaki \& Murakomi \cite{KanzakiMurakomi15}) and while such constraints on CO$_2$ for the mid Archean and earlier are not available, it is possible that CO$_2$ was the dominant component of the atmosphere during the Sun's highest activity phase.

The effects of having a CO$_2$ dominated atmosphere on the upper atmosphere of a planet can be seen well by comparing modern Earth and Venus. 
Venus' atmosphere is almost entirey CO$_2$ and despite receiving double the solar radiative flux as the Earth, its thermosphere is far colder and more contracted, with the exobase altitude being typically below 200~km and the exobase temperatures reaching only $\sim$200~K (Hedin et al. \cite{Hedin83}; Gilli et al. \cite{Gilli17}).
The effects of large changes in the CO$_2$ abundances in Earth's atmosphere were studied by Kulikov et al. (\cite{Kulikov07}) and Johnstone et al. (\cite{Johnstone18}) and an example of the effects of increasing the atmospheric CO$_2$ by a factor of 100 are shown in Fig.~\ref{fig:thermoearth}.
The green and blue lines in Fig.~\ref{fig:herbert} show the effects of on the exobase radius of having 100x the modern CO$_2$ and a 96\% CO$_2$ abundance respectively.
The higher CO$_2$ abundances lead to much reduced exobase altitudes that are below the upper estimates for the magnetopause radius.
The atmospheric loss rates from all important loss mechanisms in the cooler and less expanded cases are much lower.

The example of Earth's atmosphere evolution is useful for understanding the importance of atmospheric composition on the long term evolution of planetary atmospheres. 
While the influence of an intrinsic magnetic field on atmospheric losses is currently disputed and it is not clear if magnetospheres are indeed effective at protecting atmospheres from losses, it is clear that strong coolant molecules such as CO$_2$ are very effective at protecting atmospheres from interactions with stellar winds and from losses to space.   
Other important molecules for cooling include H$_2$O, NO, O, H, and H$_3^+$.
Although H$_2$O cooling is unimportant for modern Earth, this is because the cold trap prevents significant amounts of water reaching the the upper atmosphere.
Under many conditions the cold trap does not form, such as those of the Earth prior to the great oxidation event 2.5~Gyr ago (Gebauer et al. \cite{Gebauer17}) and those of Earth-like planets orbiting low-mass stars (Gebauer et al. \cite{Gebauer18}), and this means that H$_2$O could play a much more important role.
Cooling from species such as H$_3^+$ (see Miller et al. \cite{Miller13}) and H can be especially important for hydrogen dominated atmospheres such as those of gas giants (Murray-Clay et al. \cite{MurrayClay09}; Shaikhislamov et al. \cite{Shaikhislamov14}; Chadney et al. \cite{Chadney16}).
Note that not all cooling involves infrared emission: for example, hydrogen atoms cool by emitting in the Ly-$\alpha$ line which is part of the far ultraviolet.

\subsection{The importance of the star's rotational evolution}

\begin{figure}
\centering
\includegraphics[width=0.7\textwidth]{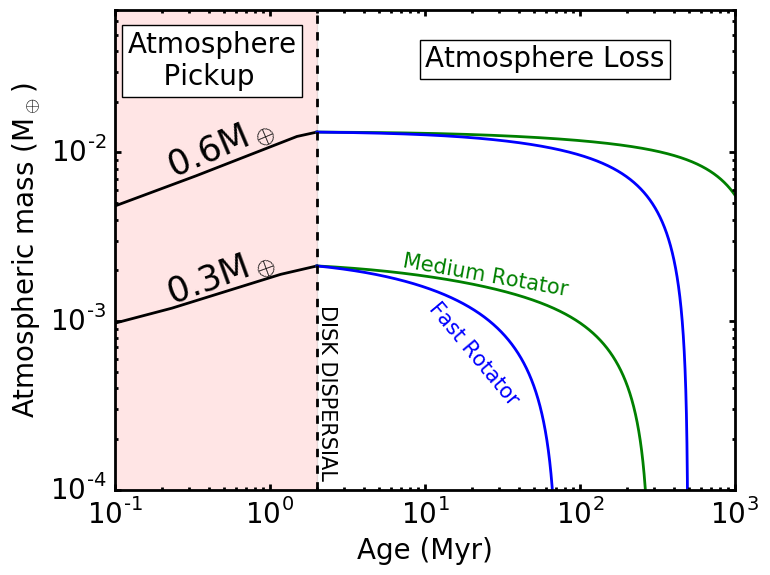}
\vskip -0mm
\caption{
\emph{
Example evolutionary tracks for the masses of hydrogen dominated primordial atmospheres on planets with masses of 0.3 and 0.6~M$_\oplus$ orbiting solar mass stars at 1~AU.
Cases for initially slowly and rapidly rotating stars are considered using the XUV evolution tracks of Tu et al. (\cite{Tu15}).
The pickup of the atmosphere during the disk phase is shown in the red shaded area and is based on the result of St\"okl et al. (\cite{Stoekl16}) and the loss after the disk has dissipated is calculated using the model developed by Johnstone et al. (\cite{Johnstone15letter}).
This shows the importance of a star's initial rotation rate for an atmosphere's evolution.
While it is assumed in the calculations that the planets remain at the masses indicated for their entire lives, what matters most is the mass of the planet at the end of the gas disk phase.
}
}
\label{fig:primattracks}
\end{figure}

Another important factor that can change the scenario for the Earth's thermosphere evolution shown in Fig.~\ref{fig:herbert} is the rotation history of the Sun.
Lichtenegger et al. (\cite{Lichtenegger10}) assumed a single time dependent decay law for the Sun's EUV emission based on the results of Ribas et al. (\cite{Ribas05}) to estimate the evolution of the thermospheric temperature.
However, as we show in Fig.~\ref{fig:rotxuvattracks}, the Sun's activity evolution depends on its initial rotation rate and different evolutionary tracks are possible.
If the Sun was born as a slow rotator, its high-energy emission during the early phases of its evolution would have been lower than that assumed by Lichtenegger et al. (\cite{Lichtenegger10}).
It is likely also that the solar wind densities and speeds would have also been lower than assumed (Johnstone et al. \cite{Johnstone15b}), though this is more speculative given our lack of understanding of the solar wind's evolution. 
Note that this section focuses on solar mass stars only and the differences between different rotation tracks is smaller for lower mass stars given their lower saturation thresholds for activity.

Given the link between stellar rotation and activity described in Section~\ref{sect:rotactivity} and the diverse ways that rotation can evolve, the rotational evolution of the central star is an important factor that influences the long term evolution of atmospheres.
This was explored for XUV driven hydrodynamic escape of primordial atmospheres in Johnstone et al. (\cite{Johnstone15letter}) and Kubyshkina et al. (\cite{Kubyshkina19})  who showed that planets orbiting initially rapidly rotating stars lose far more atmospheric gas than those orbiting initially slowly rotating stars, as demonstrated in Fig.~\ref{fig:primattracks}.
Depending on the initial rotation rate of the star, a planet can either lose an atmosphere entirely in the first billion years or keep it for its entire life, which has sever consequences for the conditions on the surfaces of these planets.
The influence of different rotation tracks was studied for Earth-mass planets with water vapor atmospheres orbiting solar mass stars at 1~AU by Johnstone et al. (\cite{Johnstone20}) who showed that initially rapidly rotating stars erode far more atmospheric gas than slowly rotating stars.
In the first billion years, a rapid rotator can remove several tens of Earth oceans of water (assuming a sufficiently large reservoir of water exists to be removed) whereas a slowly rotating star can only remove approximately 1 Earth ocean. 
Since water vapor is first photodissociated into atomic O and H (and ionized into O$^+$ and H$^+$) and the relative losses of H and O depend on the loss rate, this has significant effects on how much O$_2$ remains on the atmosphere at the end of the escape phase and on the final atmosphere's composition.

The cases shown in Fig.~\ref{fig:primattracks} end at an age of 1~Gyr and we expect little additional loss at later ages since the activities of solar mass stars decay rapidly due to stellar wind driven spin-down. 
Even in the first billion years, the total amount of atmospheric escape is reduced by the decay in stellar rotation and activity as shown by Johnstone et al. (\cite{Johnstone20}) who considered also the effects of stars remaining at their maximum activity on losses from water vapor atmospheres.
This is probably the most important effect that stellar winds have on the evolution of planetary atmospheres.
By removing angular momentum from rapidly rotating young stars, winds prevent them from remaining highly active for their entire lives and therefore prevent the early period of extreme atmospheric escape for lasting until no volatile species are left to form an atmosphere.
It is interesting to consider that despite driving a large number of important atmospheric loss processes, by spinning down their stars the net effect of stellar winds on the evolution of planetary atmospheres is likely one of atmospheric protection.

\begin{figure}
\centering
\includegraphics[width=0.85\textwidth]{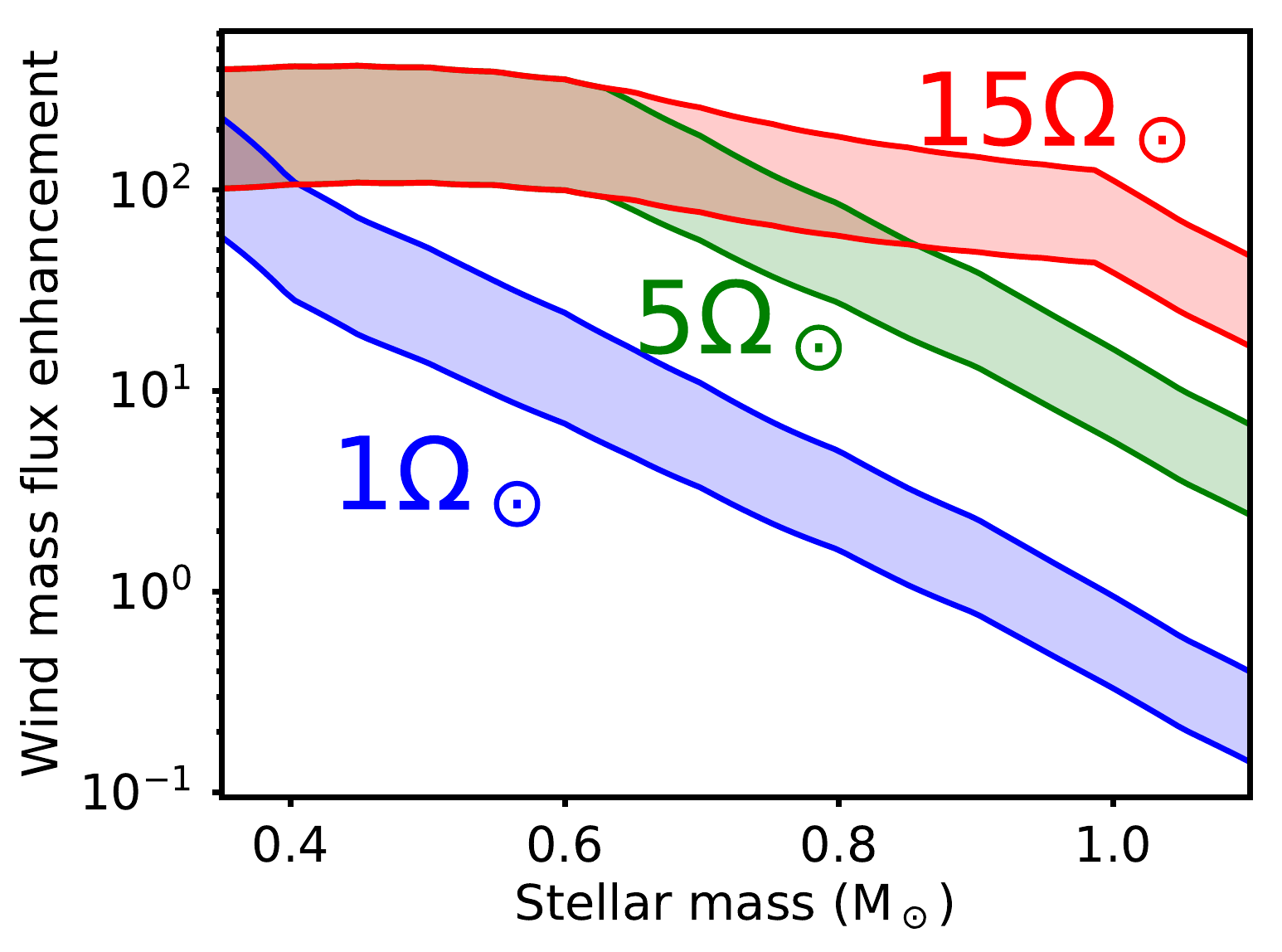}
\vskip -0mm
\caption{
\emph{
Wind mass flux in the habitable zones of main-sequence stars as a function of stellar mass for three different rotation rates.
The values are normalised to the mass flux of the solar wind at 1~AU and the ranges of values shown for each rotation rate show the fluxes between inner and outer habitable zone limits, defined as the moist and maximum greenhouse limits and calculated using the relations of Kopparapu et al. (\cite{Kopparapu13}).
}
}
\label{fig:hzwind}
\vskip -0mm
\end{figure}

\subsection{Stellar winds, CMEs, and habitable zone planets}

Much of the motivation for wanting to understand how stellar winds influence the atmospheres of planets is our desire to understand questions related to planetary habitability and the likelihood that Earth-like surface conditions form on terrestrial planets outside our solar system.
It is interesting for this purpose to consider the properties of stellar winds in the habitable zones (HZs) of stars with different masses and the effects on atmospheric losses.
In Fig.~\ref{fig:hzwind}, I show how the mass flux in the wind in the HZ depends on stellar mass for three different rotation rates which I calculate using the constraints on wind mass loss derived from rotational evolution (Eqn.~\ref{eqn:MdotRossby}) as discussed in Section~\ref{sect:windobsrot}.
Due to their much closer HZs, planets orbiting lower mass stars experience far higher mass fluxes when the stars are moderately or slowly rotating. 
Since lower mass stars experience longer periods of rapid rotation, we might expect that their HZ planets lose more atmospheric gas over longer time periods.

An important location in magnetised stellar winds is the Alfv\'en surface, which is the surface around the star at which the outflow speed of the wind becomes equal to the speed of Alfv\'en waves.
The Alfv\'en surface is typically within a few tens of stellar radii and very non-spherical.
Cohen et al. (\cite{Cohen14}) showed that for very low mass stars with HZs that are much closer to the stellar surface, the Alfv\'en surface can extend beyond the limits of the HZ, meaning that HZ planets experience both sub- and super-Alfv\'enic conditions over their orbits. 
This has an important effect on the magnetospheres of these planets since in the sub-Alfv\'enic regime, bow shows do not form and the structure of the magnetosphere is very different. 
They also found that the extreme wind conditions experienced by these planets lead to very strong Joule heating of the atmospheres, which likely contributes to the expansion and loss of the atmosphere. 
 
Interesting also is to consider the effects of CMEs which drive a large number of processes, and can even influence the lower atmospheres and climates of planets orbiting active stars.
The effects of CMEs on HZ planets orbiting low mass stars was studied by Khodachenko et al. (\cite{Khodachenko07}) and Lammer et al. (\cite{Lammer07}) who argued that planets orbiting M~dwarfs could by exposed to a continuous flux of CMEs which would compress their magnetospheres and drive a range of atmospheric loss processes. 
For such close HZs, CMEs often do not expand significantly before reaching the planet, which is also important for their influences on magnetospheres (Cohen et al. \cite{Cohenetal11}).
Considering the case of the early Earth, Airapetian et al. (\cite{Airapetian16}) found that high-energy particles accelerated as a result of a large CME that propagate downwards through the atmosphere can significantly influence the chemical composition of the gas.
The effect of a CME on loss processes in the exospheres of terrestrial planets was modelled by Kislyakova et al. (\cite{Kislyakova13}) who found stronger ionosation and pick-up of exospheric particles as CMEs pass the planet.
Most importantly, these particles have sufficient energy to break the strong triple bonds of N$_2$ molecules, and the resulting dissociation products can form molecules such as N$_2$O, which is a strong greenhouse gas and might have influenced early Earth's climate.

\section{Discussion}

In this review, I try to cover many of the aspects of stellar winds and planets that I consider important for the long term evolution of atmospheres.
Discussions of this topic often concentrate on the interactions between winds and magnetospheres and the ability of planetary magnetic fields to protect atmospheres from losses. 
This often ignores many important processes especially in planetary thermospheres where stellar X-ray and ultraviolet radiation causes heating and expansion. 
Factors such as the XUV spectrum of the star and the chemical composition of the atmosphere are potentially more important for determining atmosphere--wind interactions than the properties of the wind and the strength of the planet's magnetic field.
By cooling the upper atmosphere, chemical species such as carbon dioxide are very effective at protecting atmospheres from losses.
It is also not clear what role planetary magnetic fields play in determining losses since they both weaken some loss mechanisms while strengthening others. 
While most loss mechanisms are partly driven by winds, the net effect of winds is probably one of atmospheric protection by spinning stars down and cause their activity levels to decay over evolutionary timescales. 
A full description of these processes studied in the context of stellar activity evolution is needed to understand the influences of stellar winds on planetary atmospheres.

%%-----------------------------
%%      your bibliography
%%-----------------------------

\bibliographystyle{astron}
\bibliography{mybib}

\end{document}